# Lattices of Logical Fragments over Words

Manfred Kufleitner[*]    Alexander Lauser[*]

University of Stuttgart, FMI

**Abstract.** This paper introduces an abstract notion of fragments of monadic second-order logic. This concept is based on purely syntactic closure properties. We show that over finite words, every logical fragment defines a lattice of languages with certain closure properties. Among these closure properties are residuals and inverse $\mathcal{C}$-morphisms. Here, depending on certain closure properties of the fragment, $\mathcal{C}$ is the family of arbitrary, non-erasing, length-preserving, length-multiplying, or length-reducing morphisms. In particular, definability in a certain fragment can often be characterized in terms of the syntactic morphism. This work extends a result of Straubing in which he investigated certain restrictions of first-order logic formulae. In contrast to Straubing's model-theoretic approach, our notion of a logical fragment is purely syntactic and it does not rely on Ehrenfeucht-Fraïssé games.

As motivating examples, we present (1) a fragment which captures the stutter-invariant part of piecewise-testable languages and (2) an acyclic fragment of $\Sigma_2$. As it turns out, the latter has the same expressive power as two-variable first-order logic $FO^2$.

## 1. Introduction

A famous result of Büchi, Elgot, and Trakhtenbrot states that a language of finite words is regular if and only if it is definable in monadic second-order logic [1, 7, 24]. Later McNaughton and Papert considered first-order logic. They showed that a language is definable in first-order logic if and only if it is star-free [11]. It turned out that the class of first-order definable languages has a huge number of other characterizations; *cf.* [4]. Intuitively, a first-order definable language is easier to describe than a language which is not first-order definable. This leads to a natural notion of descriptive complexity inside the class of regular languages: The simpler the formula to describe a language, the simpler the language. Pursuing this approach, there are several possible restrictions for formulae which come to mind. For example, one can restrict the quantifier depth, the alternation depth, the number of variables, the set of atomic predicates, or the set of quantifiers, just to name a few. There are several problems connected with this approach towards descriptive complexity inside the class of regular languages. Firstly, simplicity of logical formulae is of course not a linear measure. And secondly, how do we test whether some language is definable in a given (infinite) class of formulae.

---

[*]Supported by the German Research Foundation (DFG) under grant DI 435/5-1.

There is no general solution to the first problem. Nevertheless, in some cases one can compare the expressive power of classes of formulae. Trivially, if a class of formulae is contained in another class of formulae, then we also have containment for the respective classes of languages. In some cases however, surprising inclusions and equivalences between syntactically incomparable fragments are known. For example, Thérien and Wilke have shown that a language is definable in first-order logic $FO^2$ with only two different names for the variables if and only if it is definable with one quantifier alternation [22]. Note that this is a natural restriction since first-order logic with three variables already has the full expressive power of first-order logic [9].

The solution to the second problem is usually obtained by an effective algebraic characterization. Schützenberger has shown that a language is star-free if and only if its syntactic monoid is aperiodic [15]. Together with the result of McNaughton and Papert, this yields an effective characterization of definability in first-order logic, *i.e.*, for a given regular language one can check whether this language is definable in first-order logic. This kind of correspondence between languages and finite monoids is formalized in Eilenberg's Variety Theorem [6], *e.g.*, star-free languages correspond to finite aperiodic monoids. The main idea is the following. If a class of languages $\mathcal{V}$ has certain closure properties, then there exists a class of finite monoids **V** such that a language is in $\mathcal{V}$ if and only if its syntactic monoid is in **V**. Now, if membership in **V** is decidable, then membership in $\mathcal{V}$ becomes decidable because the syntactic monoid can be computed effectively. The closure properties required by Eilenberg's Variety Theorem are Boolean operations, residuals, and inverse morphisms. A class of languages with these closure properties is called a *variety*. There are several variants and extensions of this approach. Pin has shown that there is an Eilenberg correspondence between positive varieties and ordered monoids [12]. A *positive variety* is a class of languages closed under positive Boolean operations, residuals, and inverse morphisms. A $\mathcal{C}$-variety (for some class of morphisms $\mathcal{C}$) is a class of languages closed under Boolean operations, residuals, and inverse $\mathcal{C}$-morphisms. Straubing has given an Eilenberg correspondence between $\mathcal{C}$-varieties and so-called *stamps* [18]. Here, decidability results usually rely on the syntactic morphism and not solely on the syntactic monoid. The work of Straubing was later amended by other results such as equational theories, positive $\mathcal{C}$-varieties and a wreath product for $\mathcal{C}$-varieties [2, 10, 14]. The most extensive generalization of Eilenberg's Variety Theorem is due to Gehrke, Grigorieff, and Pin [8]. They have shown that so-called lattices of languages admit an equational description. A *lattice* is a class of languages closed under positive Boolean operations. Depending on other closure properties such as residuals, the equational description of a lattice can be tightened.

So, in order to apply the existing algebraic framework to a class of languages defined by some class of formulae, it is important that the resulting class of languages has closure properties like (positive) Boolean operations, residuals, and inverse ($\mathcal{C}$-)morphisms. In this paper, we introduce a formal notion of *logical fragment* such that language classes defined by fragments admit such closure properties. In addition, almost all logical fragments in the literature also form fragments in the sense of this paper. We have chosen monadic second-order logic over words on a broad base of atoms as framework of formal logic because this setting exhausts most variants of first-order logic and monadic second-order logic found in the literature, *cf.* [3, 5, 13, 17, 18, 19, 20, 21, 23]. It includes atomic predicates for order, successor, and modular predicates as well as quantifiers for first-order, second-order and modular counting quantification.

The usual approach to closure properties of logical fragments is either indirect (*i.e.*, by showing equivalence with some class of languages for which the closure properties are already known) or it relies on methods such as Ehrenfeucht-Fraïssé games; see *e.g.* [17]. The most general result along this model-theoretic line of work is due to Straubing [18, Theorem 3]. For several combinations of restrictions within first-order logic (quantifier depth, number of variables, numerical predicates, and set of quantifiers), he showed closure under residuals and inverse $\mathcal{C}$-morphisms. One can obtain Straubing's result from our main Theorem 1. Sometimes, model-theoretic methods are difficult to apply to modalities such as modular quantifiers. Our syntactic transformation in contrast, allows to treat modular quantifiers uniformly as one of many cases of how to compose formulae. Moreover, it is conceptually easy to extend fragments by modalities which we did not consider in this paper.

Finally, we consider two examples which illustrate the formal notion of *fragments* introduced in this paper. Both examples cannot be easily described using Ehrenfeucht-Fraïssé games. The first example is $\mathbb{B}\Sigma_1[\leq]$, *i.e.*, Boolean combinations of positive existential first-order formulae using only $\leq$. This leads to stutter-invariant piecewise testable languages. The second example is a "syntactic" fragment of $\Sigma_2$ which is expressively complete for two-variable first-order logic $\text{FO}^2$. This restriction of $\Sigma_2$ requires that the comparison graph be acyclic. The vertices of the comparison graph are the variables and the edges reflect the comparisons. The resulting characterization of $\text{FO}^2$ is complementary to the result of Thérien and Wilke who showed that $\text{FO}^2$ and the "semantic" fragment $\Delta_2$ of $\Sigma_2$ have the same expressive power [22].

For a concise presentation of the main results, most proofs were moved to the appendix.

## 2. Preliminaries

A *language* over an alphabet $A$ is a subset of finite words in $A^*$. The empty word is 1 and $A^+ = A^* \setminus \{1\}$ is the set of finite nonempty words over $A$. The set $A^*$ of finite words over $A$ is the free monoid generated by $A$. It is *finitely generated* if $A$ is a finite set. A *residual* of a language $L \subseteq A^*$ is a language of the form $u^{-1}Lv^{-1} = \{w \in A^* \mid uwv \in L\}$ where $u, v \in A^*$. It is a *left residual* if $v = 1$ and it is a *right residual* if $u = 1$. Let $h : B^* \to A^*$ be a morphism between free monoids. The *inverse image* $h^{-1}(L)$ of $L$ under $h$ is the language $h^{-1}(L) = \{w \in B^* \mid h(w) \in L\}$ over $B$. The morphism $h$ is *non-erasing* (respectively, *length-reducing*, *length-preserving*) if for all $b \in B$ we have $h(b) \in A^+$ (respectively, $h(b) \in A \cup \{1\}$, $h(b) \in A$). The morphism $h$ is *length-multiplying* if there exists $m \in \mathbb{N}$ such that $h(b) \in A^m$ for all $b \in B$. Note that by the universal property of free monoids, a morphism between free monoids is completely determined by its images of the letters. If $\mathcal{C}$ is a family of morphisms, then $h$ is a $\mathcal{C}$-*morphism* if $h \in \mathcal{C}$. We introduce the following families of morphisms: all morphisms $\mathcal{C}_{all}$ between finitely generated free monoids, non-erasing morphisms $\mathcal{C}_{ne}$, length-multiplying morphisms $\mathcal{C}_{lm}$, length-reducing morphisms $\mathcal{C}_{lr}$, length-preserving morphisms $\mathcal{C}_{lp}$.

**Logic over Words.** We consider *monadic second-order logic* interpreted over finite words. In the context of logic, words are viewed as labeled linear orders. Positions are positive integers with 1 being the first position. Labels come from a fixed countable universe of letters $\Lambda$. The set of variables is $\mathbb{V}_1 \mathbin{\dot{\cup}} \mathbb{V}_2$ where $\mathbb{V}_1$ is an infinite set of first-order variables and $\mathbb{V}_2$ is an infinite set of second-order variables. First-order variables range over positions of the word and are denoted by lowercase letters (*e.g.*, $x, y, x_i \in \mathbb{V}_1$) whereas second-order variables range over subsets of positions and are denoted by uppercase letters (*e.g.*, $X, Y, X_i \in \mathbb{V}_2$). Atomic formulae include

- the constants $\top$ (for true) and $\bot$ (for false),
- the 0-ary predicate "empty" which is only true for the empty model,
- the label predicate $\lambda(x) = a$ holds if the position of $x$ is labeled with $a \in \Lambda$,
- the second-order predicate $x \in X$ which is true if $x$ is contained in $X$

and the following numerical predicates:

- the first-order equality predicate $x = y$,
- the (strict and non-strict) order predicates $x < y$ and $x \leq y$,
- the successor predicate $\mathrm{suc}(x, y)$ with the interpretation $x + 1 = y$,
- the minimum and maximum predicate $\min(x)$ and $\max(x)$ which identify the first and the last position, respectively,
- the modular predicate $x \equiv r \pmod{q}$ which is true if the position of $x$ is congruent to $r$ modulo $q$.

Formulae $\varphi$ and $\psi$ can be composed by the Boolean connectives (*i.e.*, by negation $\neg \varphi$, disjunction $\varphi \vee \psi$, and conjunction $\varphi \wedge \psi$), by existential and universal first-order quantification $\exists x\, \varphi$ and $\forall x\, \varphi$, by existential and universal second-order quantification $\exists X\, \varphi$ and $\forall X\, \varphi$, and by modular counting quantification $\exists^{r \bmod q} x\, \varphi$. The latter is true if, modulo $q$, there are $r$ positions for $x$ which make $\varphi$ true. Parentheses may be used for disambiguation and to increase readability. The set $\mathrm{FV}(\varphi) \subseteq \mathbb{V}_1 \cup \mathbb{V}_2$ of *free variables* of $\varphi$ is defined as usual. A *sentence* is a formula without free variables.

We only give a sketch of the *formal semantics* of formulae. A precise definition can be found in Appendix A. In the course of the evaluation of a formula, it is necessary to handle formulae with free variables. The idea is to encode their interpretation by enlarging the alphabet to include sets of variables. A first-order variable evaluates to a position $i$ if the label of $i$ contains the name of this variable. Similarly, a position $i$ is contained in the evaluation of a second-order variable if the variable name is contained in the label of $i$. Specifically, if $\varphi$ is a formula and $V$ is a set of variables such that $\mathrm{FV}(\varphi) \subseteq V$, then the semantics $[\![\varphi]\!]_V$ is a set of words $w = (a_1, J_1) \cdots (a_n, J_n)$ with $a_i \in \Lambda$ and $J_i \subseteq V$ such that for every first-order variable $x$ there exists exactly one position $i$ such that $x \in J_i$. The *interpretation* of a free first-order variable $x$ is then given by $x(w) = i$ for this unique index. For a second-order variable $X$ the interpretation $X(w) = \{i \in \{1, \ldots, n\} \mid X \in J_i\}$ is the set of positions containing $X$. With this, it is straightforward to define the semantics so as to coincide with the intuition given above.

We define the following particular classes of formulae:

- MSO$\mathrm{MOD}$ is the class of all formulae.
- MSO is the class of all formulae without the quantifier $\exists^{r \bmod q}$.
- FO$\mathrm{MOD}$ is the class of all first-order formulae including modular quantifiers (*i.e.*, without second-order variables).
- FO is the class of all formulae in FO$\mathrm{MOD}$ without the quantifier $\exists^{r \bmod q}$.

Let $\mathcal{F}$ be a class of formulae. For a set $\mathcal{P} \subseteq \{\mathrm{empty}, <, \leq, =, \mathrm{suc}, \min, \max, \equiv\}$ of predicates denote by $\mathcal{F}[\mathcal{P}]$ the class of formulae in $\mathcal{F}$ which (apart from $\top$, $\bot$, labels, and atomic formulae of the form $x \in X$) only use predicates in $\mathcal{P}$. This notation is refined by the following. For a class of $\mathcal{P}$ of atomic formulae let $\mathcal{F}[\mathcal{P}]$ be the class of formulae in $\mathcal{F}$ which (apart from $\top$, $\bot$, labels, and atomic formulae of the form $x \in X$) only uses atomic formulae in $\mathcal{P}$. For example, FO$[<]$ consists of all first-order formulae which only use atomic formulae of the form $\top$, $\bot$, $\lambda(x) = a$ and $x < y$ for arbitrary $x, y \in \mathbb{V}_1$, whereas $\mathrm{FO}[x_1 < x_2]$ only allows atomic formulae $\top$, $\bot$, $\lambda(x) = a$ and $x_1 < x_2$, but no order comparison of any other first-order variables. We also say that $\mathcal{F}$ *contains* some specific predicate (like the successor predicate) if there is a formula in $\mathcal{F}$ which uses this predicate.

**Logic and Languages.** For a sentence $\varphi$ and an alphabet $A \subseteq \Lambda$ the *language (over $A$) defined by $\varphi$* is the set $L_A(\varphi) = \{a_1 \cdots a_n \mid (a_1, \emptyset) \cdots (a_n, \emptyset) \in [\![\varphi]\!]_{A,\emptyset}\}$. It is the projection onto the letter-component of $[\![\varphi]\!]_{A,\emptyset}$. If the alphabet is clear from the context, then we drop it from the subscript and write $L(\varphi)$. For a class of formulae $\mathcal{F}$ the *class of languages $\mathcal{L}(\mathcal{F})$ defined by $\mathcal{F}$* maps every finite alphabet $A \subseteq \Lambda$ to the set

$$\mathcal{L}_A(\mathcal{F}) = \{L_A(\varphi) \mid \varphi \in \mathcal{F} \text{ is a sentence}\}$$

of languages over $A$. For a class of languages $\mathcal{G}$ the *class of languages defined by $\mathcal{F}$ over $\mathcal{G}$* is the class of languages mapping $A$ to $\mathcal{L}_A(\mathcal{F}) \cap \mathcal{G}_A$. Specifically, the *class of languages defined by $\mathcal{F}$ over nonempty words* maps $A$ to $\mathcal{L}_A(\mathcal{F}) \cap A^+$. Note that in $L_A(\varphi)$, the alphabet $A$ and the set of labels used in a formula $\varphi$ may well be incomparable; a label predicate $\lambda(x) = a$ with $a \notin A$ will always be false when considering the semantics over $A$. On the other hand, a formula need of course not use all labels of the alphabet over which structures are built. For example, consider the formula $\exists x \colon \lambda(x) = a$ requiring that there be an $a$-position. If $a \notin A$, then $L_A(\varphi) = \emptyset$ because all positions of a word $w$ over $A$ are non-$a$-positions; interpreted over the alphabet $A = \{a, b\}$ this formula defines the language $A^*aA^*$. This might seem unintuitive at first glance but allows a more uniform handling of languages over different alphabets and avoids tedious notation and many case-distinctions.

**Fragments.** In this section fragments are introduced as classes of formulae with natural closure properties on the syntactic level. As we shall see in Section 3, these syntactic properties transfer to closure under natural semantic operations.

A *context* is a formula with a unique occurrence of an additional constant predicate $\circ$ (to be read as "hole"). It is *primitive* if it does not use any label predicate. We shall denote primitive contexts by $\mu$ and contexts that *a priori* need not be primitive by $\nu$. The intuition is that $\circ$ is a place-holder where a formula can be plugged in. Let $\nu(\varphi)$ be the result of substituting $\varphi$ for the unique occurrence of $\circ$ in $\nu$. Contexts allow to elegantly describe subformulae as $\varphi$ is a subformula of $\psi$ if and only if there exists a context $\nu$ such that $\psi = \nu(\varphi)$.

**Definition 1** *A* fragment $\mathcal{F}$ *is a nonempty class of formulae such that for all primitive contexts $\mu$, all formulae $\varphi, \psi$, all $a \in \Lambda$ and all $x, y \in \mathbb{V}_1$:*

1. *If $\mu(\varphi) \in \mathcal{F}$, then $\mu(\top) \in \mathcal{F}$ and $\mu(\bot) \in \mathcal{F}$ and $\mu\big(\lambda(x) = a\big) \in \mathcal{F}$,*
2. *$\mu(\varphi \vee \psi) \in \mathcal{F}$ if and only if $\mu(\varphi) \in \mathcal{F}$ and $\mu(\psi) \in \mathcal{F}$,*
3. *$\mu(\varphi \wedge \psi) \in \mathcal{F}$ if and only if $\mu(\varphi) \in \mathcal{F}$ and $\mu(\psi) \in \mathcal{F}$,*
4. *if $\mu(\exists x \, \varphi) \in \mathcal{F}$ and $x \notin \mathrm{FV}(\varphi)$, then $\mu(\varphi) \in \mathcal{F}$.*

*It is* closed under negation *if $\varphi \in \mathcal{F}$ implies $\neg \varphi \in \mathcal{F}$.* $\diamondsuit$

Next, we give an intuition for fragments in terms of local substitution operations. Let $\mathcal{F}$ be a class of formulae and let $\varphi$ and $\psi$ be formulae. The *syntactic preorder* of $\mathcal{F}$ is defined by $\varphi \leq_\mathcal{F} \psi$ if $\mu(\psi) \in \mathcal{F}$ implies $\mu(\varphi) \in \mathcal{F}$ for every primitive context $\mu$. Intuitively $\varphi \leq_\mathcal{F} \psi$ means that, with respect to $\mathcal{F}$, the formula $\varphi$ is syntactically not more "complicated" than $\psi$.

Similarly, we let $\varphi \preceq_\mathcal{F} \psi$ if $\nu(\psi) \in \mathcal{F}$ implies $\nu(\varphi) \in \mathcal{F}$ for all contexts $\nu$. The syntactic preorder allows to reformulate some of the axioms of a fragment. For example, property (1) in Definition 1 is equivalent to $\top \leq_\mathcal{F} \varphi$ and $\bot \leq_\mathcal{F} \varphi$ and $(\lambda(x) = a) \leq_\mathcal{F} \varphi$ for all formulae $\varphi$. Note that $\varphi \preceq_\mathcal{F} \psi$ implies $\varphi \leq_\mathcal{F} \psi$. The reverse is however not true for arbitrary classes of formulae. Let for example $\mathcal{F}$ consist of all formulae containing at most one label predicate. In this case, we have $(\lambda(x) = a) \leq_\mathcal{F} \top$. If $\nu$ is the context $\circ \wedge \lambda(x) = a$, then $\nu(\top) \in \mathcal{F}$ and $\nu(\lambda(x) = a) \notin \mathcal{F}$. Hence $(\lambda(x) = a) \not\preceq_\mathcal{F} \top$. For fragments on the other hand, this cannot happen because here, $\leq_\mathcal{F}$ and $\preceq_\mathcal{F}$ are equivalent by the following lemma.

**Lemma 1** *If $\mathcal{F}$ is a fragment, then $\varphi \leq_\mathcal{F} \psi$ if and only if $\varphi \preceq_\mathcal{F} \psi$.* □

This provides an intuition for fragments: In a formula from a fragment $\mathcal{F}$, one may replace arbitrary subformulae by $\leq_\mathcal{F}$-smaller formulae without leaving $\mathcal{F}$. Note that this is not immediate from the definition of a fragment because in general, primitive contexts are not sufficient to formalize subformulae (as the "rest" of the formula may contain label predicates). On the other hand it is also natural to attach an alphabet to a formula (which it is interpreted over) and in this case, primitive contexts do not interfere with the alphabet of the formula.

## 3. Fragments and $\mathcal{C}$-varieties

This section summarizes semantic closure properties of fragments. In Proposition 1 and Proposition 2 we give conditions for a fragment to be closed under residuals and inverse morphisms, respectively. The combination of these two propositions gives our main result Theorem 1 which formulates closure properties of languages defined by fragments in terms of $\mathcal{C}$-varieties. For closure under residuals we need some more assumptions.

**Definition 2** *A fragment $\mathcal{F}$ is* suc-stable *if for all primitive contexts $\mu$ and all $x, y \in \mathbb{V}_1$:*
   1. *If $\mu(\mathrm{suc}(x, y)) \in \mathcal{F}$, then $\mu(x = y) \in \mathcal{F}$.*
   2. *If $\mu(\mathrm{suc}(x, y)) \in \mathcal{F}$, then $\mu(\max(x)) \in \mathcal{F}$ and $\mu(\min(y)) \in \mathcal{F}$.*
   3. *If $\mu(\min(x)) \in \mathcal{F}$ or if $\mu(\max(x)) \in \mathcal{F}$, then $\mu(\mathrm{empty}) \in \mathcal{F}$.*

*It is* mod-stable *if for all primitive contexts $\mu$, all formulae $\varphi$, all $x \in \mathbb{V}_1$ and all $q, r \in \mathbb{Z}$:*
   1. $\mu(x \equiv r \pmod{q}) \in \mathcal{F}$ *if and only if* $\mu(x \equiv s \pmod{q}) \in \mathcal{F}$ *for all $s \in \mathbb{Z}$.*
   2. $\mu(\exists^{r \bmod q} x\, \varphi) \in \mathcal{F}$ *if and only if* $\mu(\exists^{s \bmod q} x\, \varphi) \in \mathcal{F}$ *for all $s \in \mathbb{Z}$.*
   3. *If $\mu(\exists^{r \bmod q} x\, \varphi) \in \mathcal{F}$ and $x \notin \mathrm{FV}(\varphi)$, then $\mu(\varphi) \in \mathcal{F}$ and $\mu(\neg \varphi) \in \mathcal{F}$.* ◇

Consider the left residual, *i.e.*, given a formula $\varphi$ and a word $w$, we want to determine the truth value of $\varphi$ on $aw$. Conceptually, setting a variable to the "phantom" $a$-position in front of the word is handled syntactically resulting in a formula $a^{-1}\varphi$ defining the residual. To do this consistently, we keep track of these variables using the extended alphabet from the formal semantics of formulae. The above stability properties thereby allow to sustain $a^{-1}\varphi \leq_\mathcal{F} \varphi$ as an invariant. The actual construction is rather lengthy and can be found in Appendix C.

**Proposition 1** *Let $\mathcal{F}$ be a fragment and suppose that $\mathcal{F}$ is suc-stable and mod-stable. Then the class of languages defined by $\mathcal{F}$ is closed under residuals.* □

Note that if $\mathcal{F}$ does not contain suc, max or min, then $\mathcal{F}$ trivially is suc-stable. Similarly, $\mathcal{F}$ is mod-stable if it does not contain a modular predicate.

We now turn to closure under inverse morphisms. Here, we need the following additional properties of fragments.

**Definition 3** *A fragment $\mathcal{F}$ is* order-stable *if $\mu(x < y) \in \mathcal{F}$ if and only if $\mu(x \leq y) \in \mathcal{F}$ for all primitive contexts $\mu$ and all $x, y \in \mathbb{V}_1$.*

*It is* MSO-stable *if for all primitive contexts $\mu$, all formulae $\varphi$, all $x \in \mathbb{V}_1$ and all $X, Y \in \mathbb{V}_2$:*

1. *If $\mu(x \in X) \in \mathcal{F}$, then $\mu(x \in Y) \in \mathcal{F}$.*
2. *If $\mu(\exists X\, \varphi) \in \mathcal{F}$, then $\mu(\exists Y\, \exists X\, \varphi) \in \mathcal{F}$.*
3. *If $\mu(\forall X\, \varphi) \in \mathcal{F}$, then $\mu(\forall Y\, \forall X\, \varphi) \in \mathcal{F}$.* ◇

We obtain the result as follows. For every morphism $h : B^* \to A^*$ and every formula $\varphi$ we construct a formula $h^{-1}(\varphi)$ defining the inverse morphic image of $L_A(\varphi)$ with $h^{-1}(\varphi) \leq_{\mathcal{F}} \varphi$. Basically, a position $i$ on $h(w)$ can be represented by its corresponding position on $w$ (called the origin of $i$) combined with some offset (bounded by the maximal length $|h(b)|$ for letters $b \in B$). For first-order variables the offset is stored syntactically and second-order variables are distributed over several variables, depending on the offset. As for residuals, the actual construction is technically involved and can be found in Appendix D.

Typically, if a fragment $\mathcal{F}$ contains more modalities, then either $\mathcal{F}$ has to satisfy more closure properties or it is closed under fewer inverse morphisms. This trade-off between closure properties and inverse morphisms is given by the implications in Proposition 2, each implication covering certain modalities in $\mathcal{F}$.

**Proposition 2** *Let $\mathcal{F}$ be a fragment and let $\mathcal{C}$ be a family of morphisms between finitely generated free monoids. Suppose the following:*

1. *If $\mathcal{F}$ contains a second-order quantifier, then $\mathcal{F}$ is MSO-stable or all $\mathcal{C}$-morphisms are length-reducing.*
2. *If $\mathcal{F}$ contains the predicate $\leq$ or $<$, then $\mathcal{F}$ is order-stable or all $\mathcal{C}$-morphisms are length-reducing.*
3. *If $\mathcal{F}$ contains the predicate* suc, min, max *or* empty*, then all $\mathcal{C}$-morphisms are non-erasing.*
4. *If $\mathcal{F}$ contains a modular predicate, then all $\mathcal{C}$-morphisms are length-multiplying and either $\mathcal{F}$ is mod-stable or all $\mathcal{C}$-morphisms are length-preserving.*
5. *If $\mathcal{F}$ contains a modular quantifier, then $\mathcal{F}$ is mod-stable or all $\mathcal{C}$-morphisms are length-reducing.*

*Then the class of languages defined by $\mathcal{F}$ is closed under inverse $\mathcal{C}$-morphisms.* □

In particular every fragment is closed under length-preserving morphisms.

We now turn to $\mathcal{C}$-varieties of which we only give the definition; for details see [14, 18]. A *category* $\mathcal{C}$ of morphisms between finitely generated free monoids is a family of morphisms between finitely generated free monoids which contains the identity morphisms and which is closed under composition. A *positive $\mathcal{C}$-variety* is a class of languages which is closed under positive Boolean combination, residuals and inverse $\mathcal{C}$-morphisms. It is a *$\mathcal{C}$-variety* if it is closed under complement. Examples for categories of morphisms include $\mathcal{C}_{all}$, $\mathcal{C}_{ne}$, $\mathcal{C}_{lm}$, $\mathcal{C}_{lr}$, and $\mathcal{C}_{lp}$.

Our main result is the next theorem from which in particular the main results of a paper by Straubing can be obtained [18, Theorem 3]. Intuitively, the more closure properties some fragment $\mathcal{F}$ has, the larger is the class of inverse morphisms under which $\mathcal{L}(\mathcal{F})$ is closed. In Theorem 1 below this is formalized by a sequence of implications.

**Theorem 1** Let $\mathcal{F}$ be a mod-stable and suc-stable fragment. Let $\mathcal{C}$ be a category of morphisms between finitely generated free monoids. Suppose the following:

1. If $\mathcal{F}$ contains a second-order quantifier, then $\mathcal{F}$ is MSO-stable or all $\mathcal{C}$-morphisms are length-reducing.
2. If $\mathcal{F}$ contains the predicate $\leq$ or $<$, then $\mathcal{F}$ is order-stable or all $\mathcal{C}$-morphisms are length-reducing.
3. If $\mathcal{F}$ contains the predicate suc, min, max or empty, then all $\mathcal{C}$-morphisms are non-erasing.
4. If $\mathcal{F}$ contains a modular predicate, then all $\mathcal{C}$-morphisms are length-multiplying.

Then the class of languages defined by $\mathcal{F}$ is a positive $\mathcal{C}$-variety.

*Proof:* We have to show that $\mathcal{L}(\mathcal{F})$ is closed under union, intersection, residuals and inverse $\mathcal{C}$-morphisms. Using the primitive context $\circ$ it is easy to see that $\mathcal{F}$ is closed under disjunction and conjunction and, consequently, $\mathcal{L}(\mathcal{F})$ is closed under union and intersection. It remains to show that $\mathcal{L}(\mathcal{F})$ is closed under residuals and inverse $\mathcal{C}$-morphisms. Closure under residuals is Proposition 1 and closure under inverse $\mathcal{C}$-morphisms is Proposition 2. □

A *(positive) $*$-variety* is a (positive) $\mathcal{C}_{all}$-variety and a *(positive) $+$-variety* is a (positive) $\mathcal{C}_{ne}$-variety of languages of nonempty words. We get the following corollaries for fragments using equality, order and successor. Note in particular that the predicate "empty" is void over nonempty words and that every first-order fragment trivially is MSO-stable.

**Corollary 1** Let $\mathcal{F} \subseteq \text{MSO}[<, \leq, =]$ be a fragment which is MSO-stable and order-stable. Then $\mathcal{F}$ defines a positive $*$-variety. □

**Corollary 2** Let $\mathcal{F} \subseteq \text{MSO}[<, \leq, =, \text{suc}, \min, \max]$ be an MSO-stable and order-stable fragment. Suppose $\min(y) \leq_{\mathcal{F}} \text{suc}(x, y)$ and $\max(x) \leq_{\mathcal{F}} \text{suc}(x, y)$ for all first-order variables $x$ and $y$. Then the class of languages defined by $\mathcal{F}$ over nonempty words forms a positive $+$-variety. □

## 4. Stutter-Invariant Piecewise Testable Languages

A language is a *simple monomial* if it is of the form $A^*a_1 \cdots A^*a_n A^*$. A language $L \subseteq A^*$ is *piecewise testable* if it is a finite Boolean combination of simple monomials. It is *stutter-invariant* if $paq \in L$ if and only if $paaq \in L$ for all $a \in A$.

Let $\Sigma_1$ consist of all FO-formulae without negation and without any universal quantifier. Let $\mathbb{B}\Sigma_1$ be the fragment which consists of all Boolean combinations of formulae in $\Sigma_1$. By Theorem 1, the class of languages definable in $\mathbb{B}\Sigma_1[\leq]$ forms a $\mathcal{C}_{lr}$-variety. The following proposition describes the class of languages definable in $\mathbb{B}\Sigma_1[\leq]$ in terms of stutter-invariant piecewise testable languages.

**Proposition 3** Let $L \subseteq A^*$ be a language. The following are equivalent:

1. $L$ is definable in $\mathbb{B}\Sigma_1[\leq]$.
2. $L$ is piecewise testable and stutter-invariant.
3. $L$ is a Boolean combination of simple monomials of the form $A^*a_1 \cdots A^*a_n A^*$ with $a_i \neq a_{i+1}$ for all $i$.

*Proof:* We first show "(1) ⇒ (2)". If $L$ is $\mathbb{B}\Sigma_1[\leq]$-definable, then of course $L$ is $\mathbb{B}\Sigma_1[<,=]$-definable. The latter is equivalent to $L$ being piecewise testable, see *e.g.* [5]. It is easy to see that the class of languages defined by $\Sigma_1[\leq]$ is stutter-invariant. The claim follows since stutter-invariant languages are closed under Boolean operations.

"(2) ⇒ (3)": Let $L \subseteq A^*$ be piecewise testable and stutter-invariant. Since $L$ is piecewise testable, we can write $L = \bigcup_{i=1}^s P_i \setminus \bigcup_{j=1}^t Q_j$ where $P_i$ and $Q_j$ are simple monomials. Suppose $P = (A^*a_1)^{e_1} \cdots (A^*a_n)^{e_n} A^*$ for positive integers $e_i$, $a_i \in A$ and $a_i \neq a_{i+1}$. Then $\mathrm{red}(P) = A^*a_1 \cdots A^*a_n A^*$ is the monomial obtained by discarding successive $a_i$'s with the same label. Note that $\mathrm{red}(P)$ is stutter-invariant and $P \subseteq \mathrm{red}(P)$. It suffices to show $L = \bigcup_i \mathrm{red}(P_i) \setminus \bigcup_j \mathrm{red}(Q_j)$. For the containment from left to right assume $u \in L$ and $u \in \mathrm{red}(Q_j)$ for some $j$. Let $\mathrm{red}(Q_j) = A^*a_1 \cdots A^*a_n A^*$ with $a_i \neq a_{i+1}$ and let $u = u_1 a_1 \cdots u_n a_n u_{n+1}$. Then there exist positive integers $e_i$ such that $u' = u_1 a_1^{e_1} \cdots u_n a_n^{e_n} u_{n+1} \in Q_j$. Therefore, $u' \notin L$ and, by stutter-invariance of $L$, we conclude $u \notin L$, a contradiction. For the converse let $u \in \mathrm{red}(P_i)$ for some $i$ such that $u \notin \bigcup_j \mathrm{red}(Q_j)$. Let $\mathrm{red}(P_i) = A^*a_1 \cdots A^*a_n A^*$ with $a_j \neq a_{j+1}$ and $u = u_1 a_1 \cdots u_n a_n u_{n+1}$. There exist positive integers $e_i$ such that $u' = u_1 a_1^{e_1} \cdots u_n a_n^{a_n} u_{n+1} \in P_i$ and stutter-invariance of the $\mathrm{red}(Q_j)$ yields $u' \notin \bigcup_j \mathrm{red}(Q_j)$. In particular $u' \notin \bigcup_j Q_j$ and thus $u' \in L$. By stutter-invariance of $L$ we get $u \in L$.

"(3) ⇒ (1)": Let $P = A^*a_1 \cdots A^*a_n A^*$ with $a_i \neq a_{i+1}$ for all $i$. Then $P$ is defined by the formula $\exists x_1 \cdots \exists x_n \colon \bigwedge_{i=1}^n \lambda(x_i) = a_i \wedge \bigwedge_{i=1}^{n-1} x_i \leq x_{i+1}$. Note that in this formula, $x_i \leq x_{i+1}$ implies $x_i < x_{i+1}$ since $a_i \neq a_{i+1}$. □

A famous result of Simon says that a language $L$ is piecewise testable if and only if the syntactic monoid of $L$ is finite and $\mathcal{J}$-trivial [16]. The latter property is decidable for finite monoids. Moreover, $L$ is stutter-invariant if and only if the image of every letter under the syntactic morphisms of $L$ is idempotent. Combining these observations, (2) shows that it is decidable whether a given regular language is definable in $\mathbb{B}\Sigma_1[\leq]$.

## 5. The Acyclic Fragment of $\Sigma_2$

Let $\Sigma_2$ consist of all FO-formulae without negations such that there is no path in the parse-tree with an existential quantifier after a universal quantifier, *i.e.*, on every path in the parse-tree all existential quantifiers occur before all universal quantifiers. The *comparison graph* of a formula $\varphi$ is the directed graph $G(\varphi) = (V, E)$ with $V$ being the set of variables occurring in $\varphi$ and $(x, y) \in E$ if and only if one of the atomic formulae $x < y$, $x \leq y$, $x = y$ or $y = x$ occurs in $\varphi$. It is *acyclic* if there exist no $x_1, \ldots, x_n \in V$ such that $(x_i, x_{i+1}) \in E$ and $x_n = x_1$. Note that the class of formulae in $\Sigma_2[<, \leq]$ with an acyclic comparison graph forms an order-stable fragment thus defining a positive $*$-variety. In fact, the following proposition implies that it defines a $*$-variety even though, syntactically, it is not closed under negation.

**Theorem 2** *A language is definable in $\mathrm{FO}^2[<]$ if and only if it is definable by a formula in $\Sigma_2[<, \leq]$ with an acyclic comparison graph.*

*Proof:* We only give an outline. The full proof can be found in Appendix F. The proof relies on two famous characterizations of the class of languages definable in $\mathrm{FO}^2[<]$. The first characterization is in terms of unions of unambiguous monomials and the second one is the variety **DA** of finite monoids; see [20, 5].

A language of the form $P = A_1^* a_1 \cdots A_n^* a_n A_{n+1}^*$ with $a_i \in A$ and $A_i \subseteq A$ is called a *monomial*. It is *unambiguous* if every word $u \in P$ has a unique factorization $u = u_1 a_1 \cdots u_n a_n u_{n+1}$ with $u_i \in A_i^*$. For the direction from left to right, it suffices to show that every unambiguous monomial $P = A_1^* a_1 \cdots A_n^* a_n A_{n+1}^*$ is definable by a formula in $\Sigma_2[<, \leq]$ with an acyclic comparison graph. There exists some $a_i \notin A_1 \cap A_{n+1}$ since otherwise $(a_1 \cdots a_n)^2$ would admit two different factorizations. By symmetry, we can assume $a_i \notin A_1$. For every word $u \in P$ we consider the factorization $u = q a_i r$ such that $a_i$ does not occur in the prefix $q$. Then $q$ and $r$ are contained in smaller unambiguous monomials $Q \subseteq (A \setminus \{a_i\})^*$ and $R \subseteq A^*$, respectively, such that $Q a_i R \subseteq P$. By induction, there exist formulae for $Q$ and $R$. These formulae can be combined into a formula for $P$ using so-called relativizations. The main idea here is that the position of the first $a_i$ is unique and that several variables can be used to identify this position. This allows to maintain an acyclic comparison graph.

For the converse, we show that the syntactic monoid of $L(\varphi)$ is in **DA** if $\varphi$ is in $\Sigma_2[<, \leq]$ with an acyclic comparison graph. For this, it suffices to show that for some sufficiently large integer $n \geq 1$, we have $p(uv)^n u (uv)^n q \in L(\varphi)$ if and only if $p(uv)^{3n} q \in L(\varphi)$ for all $p, q, u, v \in A^*$. It is easier to describe the outline of the proof using the terminology of Ehrenfeucht-Fraïssé games. We note that in this game, the winning condition is not defined in terms of isomorphisms of game situations and thus, it is not an Ehrenfeucht-Fraïssé game in the usual sense. Since every language definable in $\mathrm{FO}^2[<]$ is also definable in $\Sigma_2[<, \leq]$, it follows that if Spoiler starts on the word $p(uv)^{3n} q$, then Duplicator wins for arbitrary comparison graphs [22]. Hence, Spoiler starts on $p(uv)^n u (uv)^n q$. Choosing $n$ large enough, we know that after Spoiler placed his pebbles on $p(uv)^n u (uv)^n q$, there are large gaps to the left and to the right of the central factor $u$. Duplicator plays as follows: Pebbles outside the center are placed on the respective position on $p(uv)^{3n} q$. For the pebbles in the central part, Duplicators strategy basically is to make as many atomic formulae true on $p(uv)^{3n} q$ as possible. He can do this because the comparison graph is acyclic. In the second round Spoiler places his pebbles on $p(uv)^{3n} q$. Exploiting acyclicity again, Duplicator can use the gaps on $p(uv)^n u (uv)^n q$ to obtain a situation where as many atomic formulae as possible are false on $p(uv)^n u (uv)^n q$. The result is a situation such that if $x_i < x_j$ (respectively, $x_i \leq x_j$) on $p(uv)^n u (uv)^n q$ implies $x_i < x_j$ (respectively, $x_i \leq x_j$) on $p(uv)^{3n} q$. Hence, $p(uv)^n u (uv)^n q \in L$ implies $p(uv)^{3n} q \in L$. □

## 6. Conclusion

We introduced fragments as classes of formulae with natural syntactic closure properties. Among others, these syntactic closure properties yield semantic closure under positive Boolean operations for the corresponding classes of languages, *i.e.*, every fragment defines a lattice of languages. Our main result is that fragments often yield closure under residuals and inverse morphisms. These properties lead to $\mathcal{C}$-varieties, thus allowing algebraic descriptions in terms of the syntactic morphism. At the end of the paper, we considered two fragments which are not easily captured by traditional techniques such as Ehrenfeucht-Fraïssé games. The first example is the Boolean closure of $\Sigma_1[\leq]$. This fragment corresponds to the stutter-invariant subclass of piecewise testable languages. The second example is a novel characterization of the $\mathrm{FO}^2[<]$-definable languages in terms of an acyclic fragment of $\Sigma_2[<, \leq]$.

We expect our constructions to be extensible to other structures such as infinite words. Another line of work would be to assign a reasonable syntactic object to a given fragment. The hope is that this object could be used for a general framework to solve questions like "is the class of languages defined by the fragment $\mathcal{F}$ closed under complement?" or "is the class of languages defined by the fragment $\mathcal{F}$ closed under shuffle?" In classical Ehrenfeucht-Fraïssé games, a winning condition for Duplicator relies on isomorphisms between game situations. We conjecture that asymmetric winning conditions as in the proof of Theorem 2 can be used to give combinatorial counterparts for arbitrary fragments.

# Appendix

# A. Formal Syntax and Semantics of Monadic-Second Order Logic with Modular Quantifiers

Let $\Lambda$ be a countable universe of labels. The syntax of a *formula* $\varphi$ is given by

$$\top \mid \bot \mid \text{empty} \mid \lambda(x) = a \mid x \in X \mid$$
$$x = y \mid x < y \mid x \leq y \mid \text{suc}(x,y) \mid \min(x) \mid \max(x) \mid x \equiv r \pmod{q} \mid$$
$$\neg \varphi \mid \varphi_1 \vee \varphi_2 \mid \varphi_1 \wedge \varphi_2 \mid \exists x\, \varphi \mid \forall x\, \varphi \mid \exists X\, \varphi \mid \forall X\, \varphi \mid \exists^{r \bmod q} x\, \varphi$$

for $x, y \in \mathbb{V}_1$, $X \in \mathbb{V}_2$, $a \in \Lambda$, $r, q \in \mathbb{Z}$ and formulae $\varphi, \psi$. Note that we do not impose a finiteness condition on $\Lambda$ but of course for every formula only a finite subset of $\Lambda$ occurs. We stipulate the usual shortcuts $\bigvee_{i=1}^{n} \varphi_i$ and $\bigwedge_{i=1}^{n} \varphi_i$ for formulae $\varphi_i$, $i \in \{1, \ldots, n\}$, with the convention that for $n = 0$ the disjunction is $\bot$ and the conjunction is $\top$. Moreover, if $A$ is a finite subset of $\Lambda$, then $\lambda(x) \in A$ is an abbreviation for the formula $\bigvee_{a \in A} \lambda(x) = a$. Parentheses may be used for disambiguation and to increase readability.

Next, we give the formal semantics of formulae. Even though we are mainly interested in sentences, we need to handle formulae with free variables in the definition of the semantics. This is done by extending the alphabet in such a way that the interpretation of free the variables can be encoded. For a formula $\varphi$ and a set of variables $V$ containing the free variables of $\varphi$, the *semantics* is a subset of words over $\Lambda \times 2^V$ and denoted by $[\![\varphi]\!]_V$. For an alphabet $A \subseteq \Lambda$ we let $[\![\varphi]\!]_{A,V} = [\![\varphi]\!]_V \cap (A \times 2^V)^*$ be the *semantics over $A$*. The idea is that the second component allows to read of the interpretation of the free variables. Suppose a position is labeled with $(a, J)$. Then a first-order variable $x$ is at this position if and only if $x \in J$ and the second-order variable $X$ contains this position if and only if $X \in J$. Let $w = (a_1, J_1) \cdots (a_n, J_n)$ where $n \geq 0$, $a_i \in \Lambda$ and $J_i \subseteq V$. Then $w \in [\![\top]\!]_V$ if and only if for all $x \in V \cap \mathbb{V}_1$ there is exactly one index $i \in \{1, \ldots, n\}$ with $x \in J_i$. Notice that $1 \notin [\![\top]\!]_V$ if $V$ contains a first-order variable but $1 \in [\![\top]\!]_\emptyset$. We are going to define the formal semantics such that $[\![\varphi]\!]_V \subseteq [\![\top]\!]_V$. If $w \in [\![\top]\!]_V$, then the *interpretation* of $X \in V$ on $w$ is the set of positions $X(w) = \{i \in \{1, \ldots, n\} \mid X \in J_i\}$. Notice that $x(w)$ is a singleton set for every first-order variable $x$ and by abuse of notation we also write $x(w)$ for the position contained in this singleton. We extend the label function to first-order variables by setting $\lambda_w(x) = \lambda_w(x(w))$. Suppose $w \in [\![\top]\!]_V$. Let $[\![\bot]\!]_V = \emptyset$. For "empty" let $w \in [\![\text{empty}]\!]_V$ if and only if $|w| = 0$. For the label predicate let $w \in [\![\lambda(x) = a]\!]_V$ if and only if $\lambda_w(x) \in \{a\} \times 2^V$. For the containment predicate let $w \in [\![x \in X]\!]_V$ if and only if $x(w) \in X(w)$. For the predicate $\sim \in \{=, <, \leq\}$ let $w \in [\![x \sim y]\!]_V$ if and only if $x(w) \sim y(w)$. For the successor let $w \in [\![\text{suc}(x,y)]\!]_V$ if and only if $x(w) + 1 = y(w)$. Let $w \in [\![\min(x)]\!]_V$ if and only of $x(w) = 1$ and let $w \in [\![\max(x)]\!]_V$ if and only if $x(w) = |w|$. For the modular predicate let $w \in [\![x \equiv r \pmod{q}]\!]_V$ if and only if $x(w) \equiv r \pmod{q}$. Boolean combinations are given inductively by $[\![\neg \varphi]\!]_V = [\![\top]\!]_V \setminus [\![\varphi]\!]_V$ and $[\![\varphi_1 \vee \varphi_2]\!]_V = [\![\varphi_1]\!]_V \cup [\![\varphi_2]\!]_V$ and $[\![\varphi_1 \wedge \varphi_2]\!]_V = [\![\varphi_1]\!]_V \cap [\![\varphi_2]\!]_V$. For the semantics of the first-order quantifiers we need to introduce some more notation. Let $w[x/i] = (a_1, J_1') \cdots (a_n, J_n')$ with $J_i' = J_i \cup \{x\}$ and

$J'_j = J_j \setminus \{x\}$ for all $j \neq i$. Let

$$[\![\exists x\, \varphi]\!]_V = \{w \in [\![\top]\!]_V \mid w[x/i] \in [\![\varphi]\!]_{V \cup \{x\}} \text{ for some } i \in \{1, \ldots, |w|\}\}.$$
$$[\![\exists X\, \varphi]\!]_V = \{(a_1, J_1) \cdots (a_n, J_n) \in [\![\top]\!]_V \mid (a_1, J'_1) \cdots (a_n, J'_n) \in [\![\varphi]\!]_{V \cup \{X\}}$$
$$\text{for some } J'_1, \ldots, J'_n \text{ with } J_i \,\Delta\, J'_i \subseteq \{X\}\}.$$

Here $J \,\Delta\, J' = (J \setminus J') \cup (J' \setminus J)$ is the *symmetric difference*.

For a formula $\varphi$ and a first-order variable $x$ let $I_w(x, \varphi) = \{i \in \{1, \ldots, |w|\} \mid w[x/i] \in [\![\varphi]\!]_{V \cup \{x\}}\}$ be the set of positions $i$ of $w$ such that $\varphi$ holds if $x$ is on position $i$. For the modular quantifier let

$$[\![\exists^{r \bmod q} x\, \varphi]\!]_V = \{w \in [\![\top]\!]_V \mid |I_w(x, \varphi)| \equiv r \pmod{q}\}$$

For the universal quantifiers let $[\![\forall x\, \varphi]\!]_V = [\![\neg \exists x\, \neg\varphi]\!]_V$ and $[\![\forall X\, \varphi]\!]_V = [\![\neg \exists X\, \neg\varphi]\!]_V$. Note that if $\mathrm{FV}(\varphi) \not\subseteq V$, then the semantics $[\![\varphi]\!]_V$ is undefined. Also take notice that in case $q = 0$ the modular predicate degenerates to equality and the modular counting quantifier counts the exact number of positions, i.e., for $w \in [\![\top]\!]_V$ we have $w \in [\![x \equiv r \pmod{0}]\!]_V$ if and only if $x(w) = r$ and $w \in [\![\exists^{r \bmod 0} x\, \varphi]\!]_V$ if and only if $|I_w(x, \varphi)| = r$.

## B. Proof of Lemma 1

In order to prove Lemma 1, we need the following lemma which a substitution principle for fragments. It states that if some subformula is replaced by a $\leq_{\mathcal{F}}$-smaller formula, then the result is again $\leq_{\mathcal{F}}$-smaller.

**Lemma 2** *Let $\mathcal{F}$ be a fragment. If $\varphi \leq_{\mathcal{F}} \psi$, then $\nu(\varphi) \leq_{\mathcal{F}} \nu(\psi)$ for every context $\nu$.*

*Proof:* The proof is by induction on the structure of $\nu$. Let $\varphi \leq_{\mathcal{F}} \psi$ and suppose $\mu(\nu(\psi)) \in \mathcal{F}$. We want to show $\mu(\nu(\varphi)) \in \mathcal{F}$. If $\nu = \circ$, then $\mu(\nu(\psi)) = \mu(\psi)$. Hence, by definition, $\mu(\nu(\varphi)) = \mu(\varphi) \in \mathcal{F}$. If $\nu$ is another atomic formula, then $\nu(\varphi) = \nu(\psi)$ and the claim becomes trivial. Suppose that $\nu = \neg \nu'$ or $\nu = Q\, \nu'$ for some quantifier $Q \in \{\exists x, \forall x, \exists^{r \bmod q} x, \exists X, \forall X \mid x \in \mathbb{V}_1, X \in \mathbb{V}_2, r, q \in \mathbb{Z}\}$ and some context $\nu'$ and consider the primitive context $\mu' = \mu(\neg \circ)$ (respectively, $\mu' = \mu(Q \circ)$). Inductively, $\nu'(\varphi) \leq_{\mathcal{F}} \nu'(\psi)$. We have $\nu(\psi) = \mu'(\nu'(\psi)) \in \mathcal{F}$ and thus $\nu(\varphi) = \mu'(\nu'(\varphi)) \in \mathcal{F}$. Finally suppose $\nu = \nu'_1 \vee \nu'_2$ for some contexts $\nu'_i$. Axiom (2) yields $\mu(\nu'_i(\psi)) \in \mathcal{F}$. By induction $\nu'_i(\varphi) \leq_{\mathcal{F}} \nu'_i(\psi)$ and therefore, $\mu(\nu'_i(\varphi)) \in \mathcal{F}$. By the same axiom $\mu(\nu(\varphi)) = \mu(\nu'_1(\varphi) \vee \nu'_2(\varphi)) \in \mathcal{F}$. In case $\nu = \nu'_1 \wedge \nu'_2$ the claim follows analogously. □

**Lemma 1.** *If $\mathcal{F}$ is a fragment, then $\varphi \leq_{\mathcal{F}} \psi$ if and only if $\varphi \preceq_{\mathcal{F}} \psi$.*

*Proof:* The implication $\varphi \preceq_{\mathcal{F}} \psi \Rightarrow \varphi \leq_{\mathcal{F}} \psi$ being trivial, it suffices to show the reverse implication. Suppose $\nu(\psi) \in \mathcal{F}$ for a context $\nu$. Let $\mu$ be the primitive context obtained from $\nu$ by replacing all label predicates by $\top$. Repeated application of Lemma 2 shows $\mu(\psi) \leq_{\mathcal{F}} \nu(\psi)$ and hence $\mu(\psi) \in \mathcal{F}$. With $\varphi \leq_{\mathcal{F}} \psi$ we see $\mu(\varphi) \in \mathcal{F}$. Again with Lemma 2, we see $\nu(\varphi) \leq_{\mathcal{F}} \mu(\varphi)$ and hence $\nu(\varphi) \in \mathcal{F}$. □

# C. Proof of Proposition 1.

In order to proof Proposition 1, we give a construction for the formula defining the residual. We concentrate here on left residuals by letters; there are dual statements for right residuals. At the end of this section we make those statements for right residuals explicit that are not straightforward symmetric versions of the statements for left residuals. Residuals by words are obtained by repeated residuals by letters.

Intermediately we need to handle formulae with free variables, even though the latter application will be to sentences. We therefore give a more general construction for the residuals over the extended alphabet $\Lambda \times 2^V$ used to encode the value of the free variables. The intuition of the construction is as follows: Given a model $w$, we want to evaluate $\varphi$ on the word $(a, J)w$. The idea is to handle the "phantom" $(a, J)$-position in front of $w$ syntactically using the set $J$ for bookkeeping purposes.

We tried to keep the construction flexible. Due to different premises, this leads to a relatively high number of lemmas (four for the atomic formulae, one for Boolean connectives, one for first-order and second-order quantification, respectively, and one for modular quantification). However, they are all of a very similar structure. Plugging the lemmas of this section into an easy induction yields for every formula $\varphi$ another formula $(a, J)^{-1}\varphi$ such that

1. $(a, J)^{-1}\varphi \leq_{\mathcal{F}} \varphi$ for all "appropriate" fragments $\mathcal{F}$, and
2. $[\![(a, J)^{-1}\varphi]\!]_{V'} = (a, J)^{-1} [\![\varphi]\!]_V$

for all sets of variables $V$ with $J \cup \mathrm{FV}(\varphi) \subseteq V$ and $V' = V \setminus (J \cap \mathbb{V}_1)$.

The first property is of syntactic nature; it notably yields that $(a, J)^{-1}\varphi$ is in $\mathcal{F}$ whenever $\varphi$ is. Here, an "appropriate" fragment is a fragment which, depending on the predicates occurring in $\varphi$, potentially has some additional closure properties. In the lemmas below, $\mathfrak{F}$ will denote a family of appropriate fragments. Note that we have one formula for all such fragments (and not for every fragment a different formula), which is a stronger result than actually needed for the closure under left residuals of a fixed fragment.

The second property gives semantic correctness, i.e., $(a, J)^{-1}\varphi$ actually defines the left residual of $\varphi$. Note that for $(a, J)^{-1} [\![\varphi]\!]_V$ to makes sense, $V$ has to contain all free variables of $\varphi$ and all variables of $J$. Also note that since $[\![(a, J)^{-1}\varphi]\!]_{V'}$ is defined, in particular no first-order variable in $J$ can appear freely in $(a, J)^{-1}\varphi$.

The following lemmas give formulae for the left residual of languages defined by one of the atomic formulae. Lemma 3 deals with the formulae $\top$, $\bot$, empty, $\min(x)$, $\lambda(x) = b$, $x = y$, $x < y$, $x \leq y$ and $x \in X$ for which the closure properties of a fragment suffice. Lemma 4 and Lemma 5 are for the successor predicate $\mathrm{suc}(x, y)$ and for $\max(x)$, respectively. Our construction for $\mathrm{suc}(x, y)$ relies on being able to replace $\mathrm{suc}(x, y)$ by $\min(y)$ thus restricting the class of appropriate fragments. For $\max(x)$ the fragment must allow to replace $\max(x)$ by empty. Lemma 6 finally gives the construction for the modular predicate $x \equiv r \pmod{q}$ where we have to be able to change the remainder parameter $r$.

**Lemma 3** *Let $\varphi$ be one of the atomic formulae $\top$, $\bot$, $\lambda(x) = b$, $x = y$, $x < y$, $x \leq y$, empty, $\min(x)$ or $x \in X$. Then for all $a \in A$ and all sets of variables $J$ there exists a formula $(a, J)^{-1}\varphi$ which satisfies*

$$(a, J)^{-1}\varphi \leq_{\mathcal{F}} \varphi \quad \text{for all fragments } \mathcal{F} \text{ and}$$
$$[\![(a, J)^{-1}\varphi]\!]_{V'} = (a, J)^{-1} [\![\varphi]\!]_V$$

where $V$ is a set of variables with $J \cup \mathrm{FV}(\varphi) \subseteq V$ and $V' = V \setminus (J \cap \mathbb{V}_1)$.

*Proof:* In the following $w$ denotes some word over $\Lambda \times 2^{V'}$. For the formulae $\top$, $\bot$ and empty we let $(a,J)^{-1}\top = \top$ and $(a,J)^{-1}\bot = \bot$ and $(a,J)^{-1}\mathrm{empty} = \bot$. Note that there exists no $w$ such that $(a,J)w$ is empty which establishes the correctness of $(a,J)^{-1}\mathrm{empty}$.

Suppose $x \in J$. Then $x((a,J)w) = 1$ and $\lambda_{(a,J)w}(x) = b$ if and only if $a = b$; for $x \notin J$ we have $\lambda_{(a,J)w}(x) = \lambda_w(x)$. Hence the label predicate is given by

$$(a,J)^{-1}\bigl(\lambda(x) = b\bigr) = \begin{cases} \lambda(x) = b & \text{if } x \notin J, \\ \top & \text{if } x \in J \text{ and } a = b, \\ \bot & \text{else.} \end{cases}$$

We next consider the equality predicated and the two order predicates. We have $x((a,J)w) = y((a,J)w)$ if and only if either $x,y \in J$ or $x,y \notin J$ and $x(w) = y(w)$. We have $x((a,J)w) < y((a,J)w)$ if and only if either $x \in J$ and $y \notin J$ or $x,y \notin J$ and $x(w) < y(w)$. For the non-strict order we have $x((a,J)w) \leq y((a,J)w)$ if and only if either $x \in J$ or $x,y \notin J$ and $x(w) \leq y(w)$. We therefore let

$$(a,J)^{-1}(x = y) = \begin{cases} x = y & \text{if } x \notin J \text{ and } y \notin J, \\ \top & \text{if } x \in J \text{ and } y \in J, \\ \bot & \text{else,} \end{cases}$$

$$(a,J)^{-1}(x < y) = \begin{cases} x < y & \text{if } x \notin J \text{ and } y \notin J, \\ \top & \text{if } x \in J \text{ and } y \notin J, \\ \bot & \text{else,} \end{cases}$$

$$(a,J)^{-1}(x \leq y) = \begin{cases} x \leq y & \text{if } x \notin J \text{ and } y \notin J, \\ \top & \text{if } x \in J, \\ \bot & \text{else,} \end{cases}$$

The formula $\min(x)$ is true over $(a,J)w$, i.e., $x((a,J)w) = 1$, if and only if $x \in J$. Thus

$$(a,J)^{-1}\bigl(\min(x)\bigr) = \begin{cases} \top & \text{if } x \in J, \\ \bot & \text{else.} \end{cases}$$

Finally consider the second-order predicate $x \in X$. Suppose $x \in X$. Then $x \in X((a,J)w)$ if and only if $X \in J$. For $x \notin J$ we have $x \in X((a,J)w)$ if and only if $x \in X(w)$. Therefore let

$$(a,J)^{-1}\bigl(x \in X\bigr) = \begin{cases} x \in X & \text{if } x \notin J, \\ \top & \text{if } x \in J \text{ and } X \in J, \\ \bot & \text{else.} \end{cases}$$

It is easy to see, that all these formulae satisfy the syntactic property $(a,J)^{-1}\varphi \leq_{\mathcal{F}} \varphi$ for all fragments $\mathcal{F}$. $\square$

**Lemma 4** *Consider the atomic formula $\mathrm{suc}(x,y)$ for some $x, y \in \mathbb{V}_1$. Let $\mathfrak{F}$ be a family of fragments $\mathcal{F}$ such that $\min(y) \leq_\mathcal{F} \mathrm{suc}(x,y)$. Then for all $a \in A$ and all sets of variables $J$ there exists a formula $(a,J)^{-1}\mathrm{suc}(x,y)$ which satisfies*

$$(a,J)^{-1}\mathrm{suc}(x,y) \leq_\mathcal{F} \mathrm{suc}(x,y) \quad \text{for all } \mathcal{F} \in \mathfrak{F} \text{ and}$$
$$[\![(a,J)^{-1}\mathrm{suc}(x,y)]\!]_{V'} = (a,J)^{-1}[\![\mathrm{suc}(x,y)]\!]_V$$

*where $V$ is a set of variables with $J \cup \{x,y\} \subseteq V$ and $V' = V \setminus (J \cap \mathbb{V}_1)$.*

*Proof:* Let $w$ be a word over $\Lambda \times 2^{V'}$ and suppose $x \in J$. Then $x((a,J)w) + 1 = y((a,J)w)$ if and only if $y((a,J)w) = 2$ if and only if $y \notin J$ and $y(w) = 1$. Suppose $x \notin J$. Then $x((a,J)w) + 1 = y((a,J)w)$ if and only if $y \notin J$ and $x(w) + 1 = y(w)$. Thus

$$(a,J)^{-1}\mathrm{suc}(x,y) = \begin{cases} \mathrm{suc}(x,y) & \text{if } x \notin J \text{ and } y \notin J, \\ \min(y) & \text{if } x \in J \text{ and } y \notin J, \\ \bot & \text{else.} \end{cases}$$

This formula satisfies the syntactic property $(a,J)^{-1}\mathrm{suc}(x,y) \leq_\mathcal{F} \mathrm{suc}(x,y)$ for all $\mathcal{F} \in \mathfrak{F}$. $\square$

**Lemma 5** *Consider the atomic formula $\max(x)$ for some $x \in \mathbb{V}_1$. Let $\mathfrak{F}$ be a family of fragments $\mathcal{F}$ such that $\mathrm{empty} \leq_\mathcal{F} \max(x)$. Then for all $a \in A$ and all sets of variables $J$ there exists a formula $(a,J)^{-1}\max(x)$ which satisfies*

$$(a,J)^{-1}\max(x) \leq_\mathcal{F} \max(x) \quad \text{for all } \mathcal{F} \in \mathfrak{F} \text{ and}$$
$$[\![(a,J)^{-1}\max(x)]\!]_{V'} = (a,J)^{-1}[\![\max(x)]\!]_V$$

*where $V$ is a set of variables with $J \cup \{x\} \subseteq V$ and $V' = V \setminus (J \cap \mathbb{V}_1)$.*

*Proof:* Let $w$ be a word over $\Lambda \times 2^{V'}$ and suppose $x \in J$. Then we have $x((a,J)w) = 1 = |(a,J)w|$ if and only if $|w| = 0$. For $x \notin J$ we have $x((a,J)w) = x(w) + 1$. Thus

$$(a,J)^{-1}\max(x) = \begin{cases} \max(x) & \text{if } x \notin J, \\ \mathrm{empty} & \text{else.} \end{cases}$$

This formula satisfies the syntactic property $(a,J)^{-1}\max(x) \leq_\mathcal{F} \max(x)$ for all $\mathcal{F} \in \mathfrak{F}$. $\square$

**Lemma 6** *Consider the atomic formula $x \equiv r \pmod{q}$ for some $x \in \mathbb{V}_1$ and some $q, r \in \mathbb{Z}$. Let $\mathfrak{F}$ be a family of fragments $\mathcal{F}$ such that $x \equiv r \pmod{q}$ and $x \equiv s \pmod{q}$ are $\mathcal{F}$-equivalent for all $s \in \mathbb{Z}$. Then for all $a \in A$ and all sets of variables $J$ there exists a formula $(a,J)^{-1}\big(x \equiv r \pmod{q}\big)$ satisfying*

$$(a,J)^{-1}\big(x \equiv r \pmod{q}\big) \leq_\mathcal{F} \big(x \equiv r \pmod{q}\big) \quad \text{for all } \mathcal{F} \in \mathfrak{F} \text{ and}$$
$$[\![(a,J)^{-1}\big(x \equiv r \pmod{q}\big)]\!]_{V'} = (a,J)^{-1}[\![\big(x \equiv r \pmod{q}\big)]\!]_V$$

*where $V$ is a set of variables with $J \cup \{x\} \subseteq V$ and $V' = V \setminus (J \cap \mathbb{V}_1)$.*

*Proof:* Let $w$ be a word over $\Lambda \times 2^{V'}$ and suppose $x \in J$. Then $x((a, J)w) = 1$. Suppose $x \notin J$. Then $x((a, J)w) = 1 + x(w)$ and hence $x((a, J)w) \equiv r \pmod{q}$ if and only if $x(w) \equiv r - 1 \pmod{q}$. Thus

$$(a, J)^{-1}(x \equiv r \pmod{q}) \;=\; \begin{cases} x \equiv r - 1 \pmod{q} & \text{if } x \notin J, \\ \top & \text{if } x \in J \text{ and } r \equiv 1 \pmod{q}, \\ \bot & \text{else.} \end{cases}$$

This formulae satisfies the syntactic property. $\square$

The next lemmas "lift" the construction to formulae composed by Boolean combinations and quantifiers. These lemmas state that if there are formulae defining the left residual which work for all fragments in $\mathfrak{F}$, then there exist formulae defining the left residual of the Boolean combinations (Lemma 7), the first- and the second-order quantification (Lemma 8 and Lemma 9, respectively) which also work for all fragments in $\mathfrak{F}$. Moreover, there also exists a formula defining the left residual for a formula involving the modular counting quantifier which works for all fragments in $\mathfrak{F}$ with some additional closure property (Lemma 10).

**Lemma 7** *Let $\psi$ be one of the formulae $\neg\varphi_1$ or $\varphi_1 \vee \varphi_2$ or $\varphi_1 \wedge \varphi_2$. Let $\mathfrak{F}$ be a family of fragments. Suppose for all $a \in A$ and all sets of variables $J$ there exists a formula $(a, J)^{-1}\varphi_i$, $i \in \{1, 2\}$, which satisfy*

$$(a, J)^{-1}\varphi_i \;\leq_{\mathcal{F}}\; \varphi_i \quad \text{for all } \mathcal{F} \in \mathfrak{F} \text{ and}$$
$$[\![(a, J)^{-1}\varphi_i]\!]_{V'} \;=\; (a, J)^{-1} [\![\varphi_i]\!]_V$$

*where $V$ is a set of variables $J \cup \mathrm{FV}(\varphi_i) \subseteq V$ and $V' = V \setminus (J \cap \mathbb{V}_1)$. Then for all $a \in A$ and all sets of variables $J$ there exists a formula $(a, J)^{-1}\psi$ which satisfies*

$$(a, J)^{-1}\psi \;\leq_{\mathcal{F}}\; \psi \quad \text{for all } \mathcal{F} \in \mathfrak{F} \text{ and}$$
$$[\![(a, J)^{-1}\psi]\!]_{V'} \;=\; (a, J)^{-1} [\![\psi]\!]_V$$

*where $V$ is a set of variables with $J \cup \mathrm{FV}(\psi) \subseteq V$ and $V' = V \setminus (J \cap \mathbb{V}_1)$.*

*Proof:* The constructions for positive Boolean connectives are:

$$(a, J)^{-1}(\varphi_1 \vee \varphi_2) \;=\; \big((a, J)^{-1}\varphi_1\big) \vee \big((a, J)^{-1}\varphi_2\big),$$
$$(a, J)^{-1}(\varphi_1 \wedge \varphi_2) \;=\; \big((a, J)^{-1}\varphi_1\big) \wedge \big((a, J)^{-1}\varphi_2\big).$$

Let $\mu$ be a primitive context, let $\mathcal{F} \in \mathfrak{F}$ and suppose $\mu(\varphi_1 \vee \varphi_2) \in \mathcal{F}$. Since $\mathcal{F}$ is a fragment we see $\mu(\varphi_i) \in \mathcal{F}$ (for $i \in \{1, 2\}$). By assumption $\mu((a, J)^{-1}\varphi_i) \in \mathcal{F}$ and finally $\mu\big((a, J)^{-1}\varphi_i \vee (a, J)^{-1}\varphi_i\big) \in \mathcal{F}$. This shows $(a, J)^{-1}(\varphi_1 \vee \varphi_2) \leq_{\mathcal{F}} (\varphi_1 \vee \varphi_2)$. Analogously $(a, J)^{-1}(\varphi_1 \wedge \varphi_2) \leq_{\mathcal{F}} (\varphi_1 \wedge \varphi_2)$. For the negation let

$$(a, J)^{-1}(\neg\varphi_1) \;=\; \neg\big((a, J)^{-1}\varphi_1\big),$$

Suppose $\mu(\neg\varphi_1) \in \mathcal{F}$ for some primitive context $\mu$. Thus $\mu'(\varphi_1) \in \mathcal{F}$ for the primitive context $\mu' = \mu(\neg\circ)$. Now, by assumption $\mu'((a, J)^{-1}\varphi_1) = \mu\big(\neg(a, J)^{-1}\varphi_1\big) \in \mathcal{F}$. This shows $(a, J)^{-1}(\neg\varphi_1) \leq_{\mathcal{F}} \neg\varphi_1$. The semantic correctness is easily verified for all Boolean connectives. $\square$

**Lemma 8** *Consider the formulae $\exists x\,\varphi$ and $\forall x\,\varphi$ for some $x \in \mathbb{V}_1$. Let $\mathfrak{F}$ be a family of fragments. Suppose for all $a \in A$ and all sets of variables $J$ there exists a formula $(a,J)^{-1}\varphi$ which satisfies*

$$(a,J)^{-1}\varphi \leq_{\mathcal{F}} \varphi \quad \text{for all } \mathcal{F} \in \mathfrak{F} \text{ and}$$
$$[\![(a,J)^{-1}\varphi]\!]_{V'} = (a,J)^{-1}[\![\varphi]\!]_V$$

*where $V$ is a set of variables with $J \cup \mathrm{FV}(\varphi) \subseteq V$ and $V' = V \setminus (J \cap \mathbb{V}_1)$. Then for all $a \in A$ and all sets of variables $J$ there exist formulae $(a,J)^{-1}\exists x\,\varphi$ and $(a,J)^{-1}\forall x\,\varphi$ which satisfy*

$$(a,J)^{-1}\exists x\,\varphi \leq_{\mathcal{F}} \exists x\,\varphi \quad \text{for all } \mathcal{F} \in \mathfrak{F}, \qquad [\![(a,J)^{-1}\exists x\,\varphi]\!]_{V'} = (a,J)^{-1}[\![\exists x\,\varphi]\!]_V,$$
$$(a,J)^{-1}\forall x\,\varphi \leq_{\mathcal{F}} \forall x\,\varphi \quad \text{for all } \mathcal{F} \in \mathfrak{F}, \qquad [\![(a,J)^{-1}\forall x\,\varphi]\!]_{V'} = (a,J)^{-1}[\![\forall x\,\varphi]\!]_V$$

*where $V$ is a set of variables with $J \cup \mathrm{FV}(\exists x\,\varphi) = J \cup \mathrm{FV}(\forall x\,\varphi) \subseteq V$ and $V' = V \setminus (J \cap \mathbb{V}_1)$.*

*Proof:* Let $w$ be a word over $\Lambda \times 2^{V'}$ and let $w' = (a,J)w$. We only argue for the existential quantifier; the universal quantifier is analogue. Suppose $w'[x/i] \in [\![\varphi]\!]_{V \cup \{x\}}$ for some $i$. Suppose $i = 1$ and $w'[x/i] = (a, J \cup \{x\})w''$. Note that $w''$ is essentially $w$ but $x$ is removed from all second components. Then $w' \in [\![\varphi]\!]_{V \cup \{x\}}$ is equivalent to $w'' \in [\![(a, J \cup \{x\})^{-1}\varphi]\!]_{V' \setminus \{x\}}$ by assumption. This in turn is equivalent to $w \in [\![(a, J \cup \{x\})^{-1}\varphi]\!]_{V'}$ since, in particular, $x \notin \mathrm{FV}((a, J \cup \{x\})^{-1}\varphi)$ is not a free variable of $(a, J \cup \{x\})^{-1}\varphi$ and consequently its truth value does not depend on the value of $x$.

Let $i \geq 2$ and let $w'[x/i] = (a, J \setminus \{x\})w''$. Notice $w'' = w[x/i - 1]$. By assumption $w'[x/i] \in [\![\varphi]\!]_{V \cup \{x\}}$ is equivalent to $w'' \in [\![(a, J \setminus \{x\})^{-1}\varphi]\!]_{V' \setminus \{x\}}$. These considerations allow to set (with $\varphi_1 = (a, J \cup \{x\})^{-1}\varphi$ and $\varphi_2 = (a, J \setminus \{x\})^{-1}\varphi$)

$$(a,J)^{-1}\exists x\,\varphi = \varphi_1 \vee \exists x\,\varphi_2,$$
$$(a,J)^{-1}\forall x\,\varphi = \varphi_1 \wedge \forall x\,\varphi_2.$$

We now show the syntactic property. Let $\mathcal{F} \in \mathfrak{F}$. We have $\varphi_1 \leq_{\mathcal{F}} \exists x\,\varphi_1 \leq_{\mathcal{F}} \exists x\,\varphi$ by assumption on $\varphi$; note that $x \notin \mathrm{FV}(\varphi_1)$. Together with $\exists x\,\varphi_2 \leq_{\mathcal{F}} \exists x\,\varphi$ this yields $(a,J)^{-1}\exists x\,\varphi \leq_{\mathcal{F}} \exists x\,\varphi$. The argument for the universal quantifier is analogue. □

**Lemma 9** *Consider the formulae $\exists X\,\varphi$ and $\forall X\,\varphi$ for some $X \in \mathbb{V}_2$. Let $\mathfrak{F}$ be a family of fragments. Suppose for all $a \in A$ and all sets of variables $X$ there exists a formula $(a,J)^{-1}\varphi$ which satisfies*

$$(a,J)^{-1}\varphi \leq_{\mathcal{F}} \varphi \quad \text{for all } \mathcal{F} \in \mathfrak{F} \text{ and}$$
$$[\![(a,J)^{-1}\varphi]\!]_{V'} = (a,J)^{-1}[\![\varphi]\!]_V$$

*where $V$ is a set of variables with $J \cup \mathrm{FV}(\varphi) \subseteq V$ and $V' = V \setminus (J \cap \mathbb{V}_1)$. Then for all $a \in A$ and all sets of variables $J$ there exist formulae $(a,J)^{-1}\exists X\,\varphi$ and $(a,J)^{-1}\forall X\,\varphi$ which satisfy*

$$(a,J)^{-1}\exists X\,\varphi \leq_{\mathcal{F}} \exists X\,\varphi \quad \text{for all } \mathcal{F} \in \mathfrak{F}, \qquad [\![(a,J)^{-1}\exists X\,\varphi]\!]_{V'} = (a,J)^{-1}[\![\exists X\,\varphi]\!]_V,$$
$$(a,J)^{-1}\forall X\,\varphi \leq_{\mathcal{F}} \forall X\,\varphi \quad \text{for all } \mathcal{F} \in \mathfrak{F}, \qquad [\![(a,J)^{-1}\forall X\,\varphi]\!]_{V'} = (a,J)^{-1}[\![\forall X\,\varphi]\!]_V$$

*where $V$ is a set of variables with $J \cup \mathrm{FV}(\exists X\,\varphi) = J \cup \mathrm{FV}(\forall X\,\varphi) \subseteq V$ and $V' = V \setminus (J \cap \mathbb{V}_1)$.*

*Proof:* Let $w = (a_1, J_1) \cdots (a_n, J_n)$ and let $w' = (a, J)w$ where $a_i \in \Lambda$ and $J_i \subseteq V'$. We only argue for the existential quantifier; the universal quantifier is analogue. Suppose $(a, J')(a_1, J'_1) \cdots (a_n, J'_n) \in [\![\varphi]\!]_{V \cup \{X\}}$ for some $J'$ and $J'_i$ with $J' \Delta J \subseteq \{X\}$ and $J'_i \Delta J_i \subseteq \{X\}$. Now, either $X \in J'$ or $X \notin J'$. The premise for $\varphi$ yields $(a_1, J'_1) \cdots (a_n, J'_n) \in [\![(a, J \cup \{X\})^{-1} \varphi]\!]_{V \cup \{X\}}$ in the first case (and $(a_1, J'_1) \cdots (a_n, J'_n) \in [\![(a, J \setminus \{X\})^{-1} \varphi]\!]_{V \cup \{X\}}$ in the second case, respectively). Therefore we define

$$(a, J)^{-1} \exists X \, \varphi \;=\; \exists X \big((a, J \cup \{X\})^{-1} \varphi \vee (a, J \setminus \{X\})^{-1} \varphi\big),$$
$$(a, J)^{-1} \forall X \, \varphi \;=\; \forall X \big((a, J \cup \{X\})^{-1} \varphi \wedge (a, J \setminus \{X\})^{-1} \varphi\big).$$

Next, we show the syntactic property. Let $\mathcal{F}$ be a fragment in $\mathfrak{F}$. The assumption on $\varphi$ yields $(a, J \cup \{X\})^{-1} \varphi \leq_{\mathcal{F}} \varphi$ and $(a, J \setminus \{X\})^{-1} \varphi \leq_{\mathcal{F}} \varphi$. Thus $(a, J \cup \{X\})^{-1} \varphi \vee (a, J \setminus \{X\})^{-1} \varphi \leq_{\mathcal{F}} \varphi$ and finally $(a, J)^{-1} \exists X \, \varphi \leq_{\mathcal{F}} \exists X \, \varphi$. The argument for the universal quantifier is analogue. □

The following lemma lifts the residual construction to the modular counting quantifier. It in particular applies to mod-stable fragments but is formulated slightly more general.

**Lemma 10** *Consider the formula $\exists^{r \bmod q} x \, \varphi$ for some $x \in \mathbb{V}_1$ and some $r, q \in \mathbb{Z}$. Let $\mathfrak{F}$ be a family of fragments such that the formulae $\exists^{r \bmod q} x \, \varphi$ and $\exists^{s \bmod q} x \, \varphi$ are $\mathcal{F}$-equivalent for all $s \in \mathbb{Z}$ and such that $\psi \leq_{\mathcal{F}} \exists^{r \bmod q} x \, \varphi$ and $\neg \psi \leq_{\mathcal{F}} \exists^{r \bmod q} x \, \varphi$ for all $\psi \leq_{\mathcal{F}} \varphi$ with $x \notin \mathrm{FV}(\psi)$. Suppose for all $a \in A$ and all sets of variables $J$ there exists a formula $(a, J)^{-1} \varphi$ which satisfies*

$$(a, J)^{-1} \varphi \;\leq_{\mathcal{F}}\; \varphi \quad \text{for all } \mathcal{F} \in \mathfrak{F} \text{ and}$$
$$[\![(a, J)^{-1} \varphi]\!]_{V'} \;=\; (a, J)^{-1} [\![\varphi]\!]_V$$

*where $V$ is a set of variables with $J \cup \mathrm{FV}(\varphi) \subseteq V$ and $V' = V \setminus (J \cap \mathbb{V}_1)$. Then for all $a \in A$ and all sets of variables $J$ there exists a formula $(a, J)^{-1} (\exists^{r \bmod q} x \, \varphi)$ which satisfies*

$$(a, J)^{-1} (\exists^{r \bmod q} x \, \varphi) \;\leq_{\mathcal{F}}\; \exists^{r \bmod q} x \, \varphi \quad \text{for all } \mathcal{F} \in \mathfrak{F} \text{ and}$$
$$[\![(a, J)^{-1} (\exists^{r \bmod q} x \, \varphi)]\!]_{V'} \;=\; (a, J)^{-1} [\![\exists^{r \bmod q} x \, \varphi]\!]_V$$

*where $V$ is a set of variables with $J \cup \mathrm{FV}(\exists^{r \bmod q} x \, \varphi) \subseteq V$ and $V' = V \setminus (J \cap \mathbb{V}_1)$.*

*Proof:* Let $\varphi_1 = (a, J \cup \{x\})^{-1} \varphi$ and $\varphi_2 = (a, J \setminus \{x\})^{-1} \varphi$ be the formulae from the premise. Let

$$(a, J)^{-1} \exists^{r \bmod q} x \, \varphi \;=\; \big(\varphi_1 \wedge \exists^{r-1 \bmod q} x \, \varphi_2\big) \vee \big(\neg \varphi_1 \wedge \exists^{r \bmod q} x \, \varphi_2\big).$$

Suppose we are given a model $w$. The formula realizes a straightforward case distinction: Either the first position of $(a, J)w$ is a $\varphi$-position and then the number of $\varphi$-positions in the factor $w$ has to be $r - 1$ (modulo $q$), or it is not a $\varphi$-position and then the number of $\varphi$-positions in $w$ is $r$ (modulo $q$). Here, for conciseness we say that a position $i$ of $h_\delta(w)$ is a $\varphi$-position, if $(h_\delta(w))[x/i] \in [\![\varphi]\!]_{V \cup \{x\}}$, i.e., $\varphi$ is true if $x$ is interpreted by the position $i$. Given the closure properties of a fragment $\mathcal{F} \in \mathfrak{F}$, it is easy to see that $(a, J)^{-1} \exists^{r \bmod q} x \, \varphi \leq_{\mathcal{F}} \exists^{r \bmod q} x \, \varphi$ is inherited from $\varphi_1$ and $\varphi_2$. Notice $x \notin \mathrm{FV}(\varphi_1)$. □

**Right Residuals.** There are of course dual statements providing us with formulae $\varphi(a, J)^{-1}$ defining the right residual. We shall only make those explicit where some attention has to be paid for the premises.

**Lemma 11** *Consider the atomic formula* $\mathrm{suc}(x, y)$ *for some* $x, y \in \mathbb{V}_1$. *Let* $\mathfrak{F}$ *be a family of fragments* $\mathcal{F}$ *such that* $\max(x) \leq_{\mathcal{F}} \mathrm{suc}(x, y)$. *Then for all* $a \in A$ *and all sets of variables* $J$ *there exists a formula* $\mathrm{suc}(x, y)(a, J)^{-1}$ *which satisfies*

$$\mathrm{suc}(x, y)(a, J)^{-1} \leq_{\mathcal{F}} \mathrm{suc}(x, y) \quad \text{for all } \mathcal{F} \in \mathfrak{F} \text{ and}$$
$$[\![\mathrm{suc}(x, y)(a, J)^{-1}]\!]_{V'} = [\![\mathrm{suc}(x, y)]\!]_V (a, J)^{-1}$$

*where* $V$ *is a set of variables with* $J \cup \{x, y\} \subseteq V$ *and* $V' = V \setminus (J \cap \mathbb{V}_1)$.

*Proof:* Let $w$ be a word over $\Lambda \times 2^{V'}$ and suppose $y \in J$. Then $x(w(a, J)) + 1 = y(w(a, J)) = |w(a, J)|$ if and only if $x(w) = |w|$. Suppose now $y \notin J$, then $x((a, J)w) + 1 = y((a, J)w)$ if and only if $x \notin J$ and $x(w) + 1 = y(w)$. Let thus

$$\mathrm{suc}(x, y)(a, J)^{-1} = \begin{cases} \mathrm{suc}(x, y) & \text{if } x \notin J \text{ and } y \notin J, \\ \max(x) & \text{if } x \notin J \text{ and } y \in J, \\ \bot & \text{else.} \end{cases}$$

This formula satisfies the syntactic property. $\square$

**Lemma 12** *Consider the atomic formula* $\min(x)$ *for some* $x \in \mathbb{V}_1$. *Let* $\mathfrak{F}$ *be a family of fragments* $\mathcal{F}$ *such that* $\mathrm{empty} \leq_{\mathcal{F}} \min(x)$. *Then for all* $a \in A$ *and all sets of variables* $J$ *there exists a formula* $\min(x)(a, J)^{-1}$ *which satisfies*

$$\min(x)(a, J)^{-1} \leq_{\mathcal{F}} \min(x) \quad \text{for all } \mathcal{F} \in \mathfrak{F} \text{ and}$$
$$[\![\min(x)(a, J)^{-1}]\!]_{V'} = [\![\min(x)]\!]_V (a, J)^{-1}$$

*where* $V$ *is a set of variables with* $J \cup \{x\} \subseteq V$ *and* $V' = V \setminus (J \cap \mathbb{V}_1)$.

*Proof:* Let $w$ be a word over $\Lambda \times 2^{V'}$ and suppose $x \in J$. Then $x(w(a, J)) = |w(a, J)|$ and consequently $x(w(a, J)) = 1$ if and only if $|w| = 0$. If $x \notin J$, then $x(w(a, J)) = x(w)$. Let therefore

$$(a, J)^{-1} \min(x) = \begin{cases} \min(x) & \text{if } x \notin J, \\ \mathrm{empty} & \text{else.} \end{cases}$$

This formula satisfies the syntactic property. $\square$

We are now ready to show Proposition 1.

**Proposition 1.** *Let* $\mathcal{F}$ *be a fragment and suppose that* $\mathcal{F}$ *is suc-stable and mod-stable. Then the class of languages defined by* $\mathcal{F}$ *is closed under residuals.*

*Proof:* We show closure under left residuals. By induction on the structure of $\varphi$ we see that for for all $a \in A$ and all sets of variables $J$ there exists a formula $(a, J)^{-1}\varphi$ which satisfies $(a, J)^{-1}\varphi \leq_{\mathcal{F}} \varphi$ and $[\![(a, J)^{-1}\varphi]\!]_{V'} = (a, J)^{-1} [\![\varphi]\!]_V$ where $V$ is a set of variables with $J \cup \mathrm{FV}(\varphi) \subseteq V$ and $V' = V \setminus (J \cap \mathbb{V}_1)$. For the atomic modalities $\top$, $\bot$, $\lambda(x) = a$, $x = y$, $x < y$, $x \leq y$, empty, $\min(x)$ and $x \in X$ this is Lemma 3. Let $\mathfrak{F} = \{\mathcal{F}\}$. The predicates $\mathrm{suc}(x, y)$, $\max(x)$ and $x \equiv r \pmod{q}$ are Lemma 4, Lemma 5 and Lemma 6, respectively. Note that in all cases $\mathfrak{F}$ meets the requirements of the respective lemmas. For Boolean connectives, first-order quantification, second-order quantification and modular quantification, the claim follows by induction and Lemma 7, Lemma 8, Lemma 9 and Lemma 10, respectively. Now, using the claim and setting $a^{-1}\varphi = (a, \emptyset)^{-1}\varphi$ yields $a^{-1}L_A(\varphi) = L_A(a^{-1}\varphi)$ for every finite alphabet $A \subseteq \Lambda$ and $a^{-1}\varphi \leq_{\mathcal{F}} \varphi$. In particular, if $\varphi \in \mathcal{F}$, then $a^{-1}\varphi \in \mathcal{F}$, that is, if $L \in \mathcal{L}_A(\mathcal{F})$, then also $a^{-1}L \in \mathcal{L}_A(\mathcal{F})$. Closure of $\mathcal{L}_A(\mathcal{F})$ under right residuals follows symmetrically. □

## D. Proof of Proposition 2.

In order to proof Proposition 2, we give a construction for the formula defining the inverse morphic image. More specifically, let $A, B \subseteq \Lambda$ be finite alphabets. For a morphism $h : B^* \to A^*$ and a formula $\varphi$ we construct a formula $h^{-1}(\varphi)$ which, interpreted over $w$, has the same truth value as $\varphi$ interpreted over $h(w)$. In addition, $h^{-1}(\varphi)$ meets the syntactic property of being not "more complicated" than $\varphi$ in a certain sense.

As for residuals the application will mainly be to sentences, but we intermediately need to handle formulae with free variables. We need some more notation to formulate this concisely. Let $h : B^* \to A^*$ be a morphism, let $w = b_1 \cdots b_m$ for $b_i \in B$ and suppose $h(w) = a_1 \cdots a_n$ for $a_i \in A$. Then for every $i \in \{1, \ldots, n\}$ there exist unique numbers $j \in \{1, \ldots, m\}$ and $d \in \{1, \ldots, |h(b_j)|\}$ such that $|a_1 \cdots a_i| = |h(b_1 \cdots b_{j-1})| + d$. The numbers $(j, d)$ are called $h$-*coordinates* on $w$ of the position $i$ of $h(w)$; the number $j$ basically identifies the position of $w$ where $i$ originates from and $d$ is the *offset* within the image of $b_i$. See also Figure 1 for an illustration. Note that if $B$ is finite, then $\max_{b \in B} |h(b)|$ is a well-defined upper bound for $d$.

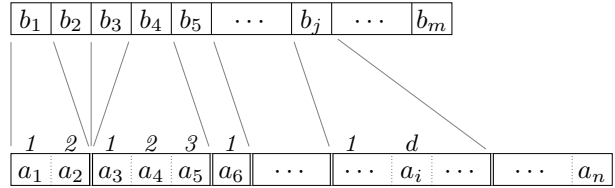

Figure 1: The $h$-coordinates of $h(w)$. In this example, $h(b_1) = a_1 a_2$, $h(b_2) = h(b_3) = 1$, $h(b_4) = a_3 a_4 a_5$ and $h(b_5) = a_6$; position 5, for example, has $h$-coordinates $(4, 3)$.

The idea is encode the variables of $h(w)$ in the alphabet of $w$. Let $i$ be a position of $h(w)$ with $h$-coordinates $(j, d)$. A first-order variable is encoded by the $h$-coordinates $(j, d)$. A second-order variable $X$ is distributed over several second-order variables $X_i$ in such a way that $X$ contains the position $i$ of $h(w)$ if $X_d$ contains the position $j$ of $w$. To formalize this we first introduce for every set of variables $V$ a derived set of variables $V_n$ with the same first-order variables as $V$ such that for every second-order variable $X$ there are $n$ distinct variables $X_1, \ldots, X_n$. If now $\delta : \mathbb{V}_1 \to_p \mathbb{N}$ is a partial function mapping a first-order variable to its offset, then the above encoding is realized by the morphism $h_\delta : (B \times 2^{V_n})^* \to (A \times 2^V)^*$ given by the following definition.

**Definition 4** *Let $V$ be a set of variables and let $n \in \mathbb{N}$. Then $V_n$ is some fixed minimal set of variables with $V_n \cap \mathbb{V}_1 = V \cap \mathbb{V}_1$ and for every second-order variable $X \in V$ we have $X \in V_n$ and there exist distinct second-order variables $X = X_1, \ldots, X_n \in V_n$ such that $\{X_1, \ldots, X_n\}$ and $\{Y_1, \ldots, Y_n\}$ are disjoint for $X \neq Y$.*

*Let $h : B^* \to A^*$ be a morphism, let $\delta : \mathbb{V}_1 \to_p \mathbb{N}$ and let $n \in \mathbb{N}$. Let the morphism $h_\delta : (B \times 2^{V_n})^* \to (A \times 2^V)^*$ be given by $h(b) = (a_1, J_1) \cdots (a_\ell, J_\ell)$ if $h(b) = a_1 \cdots a_\ell$ and*

1. *$x \in J_i$ if and only if $x \in J$ and $\delta(x) = i$ for all first-order variables $x$ and*

2. *$X \in J_i$ if and only if $X_i \in J$ for all second-order variables $X$.* ◇

Note that in order to avoid an all to tedious notation, the parameter $n$ is understood implicitly in $h_\delta$.

In this section we are going to give for every formula $\varphi$ and every "appropriate" homomorphism $h_\delta$, a formula $h_\delta^{-1}(\varphi)$ such that

1. $h_\delta^{-1}(\varphi) \leq_\mathcal{F} \varphi$ for all "appropriate" fragments $\mathcal{F}$, and
2. $[\![h_\delta^{-1}(\varphi)]\!]_{B,V_n} = h_\delta^{-1}([\![\varphi]\!]_{A,V})$

where $V$ is a set of variables with $\mathrm{FV}(\varphi) \subseteq V$.

By the first, syntactic property in particular $h_\delta^{-1}(\varphi) \in \mathcal{F}$ if $\varphi \in \mathcal{F}$. What is appropriate depends on the predicates used by the formula. The second property is semantic correctness, i.e., $h_\delta^{-1}(\varphi)$ indeed defines the inverse image of $h_\delta$; again "appropriate" depends on the predicates of the formula. Note that the statement is stronger than just closure of a fragment under inverse morphisms because we get one formula which works for all appropriate fragments. Also note that for first-order formulae we may choose $V \subseteq \mathbb{V}_1$ and thus $V_n = V$.

For the atomic formulae these are the following five lemmas. Lemma 13 gives the construction for $\top$, $\bot$, $\lambda(x) = a$ and $x = y$; Lemma 14 is for $x < y$ and $x \leq y$; Lemma 15 is for $\mathrm{suc}(x, y)$, $\min(x)$, $\max(x)$ and empty; Lemma 16 is for $x \equiv r \pmod{q}$; and Lemma 17 is for the second-order predicate $x \in X$. Subsequently, we give lifting arguments for Boolean combinations (Lemma 18), first-order, second-order and modular quantification (respectively, Lemma 19, Lemma 20 and Lemma 21).

The lemmas in this section can be seen as a toolbox for the closure under inverse morphisms of which one needs to consider only those lemmas which are relevant in a given situation. These are then connected by an easy induction. For example, for a fragment of first-order logic without modular quantifiers using only the order predicates, it suffices to consider Lemma 13 for true, false and label, Lemma 14 for the order, Lemma 18 for Boolean connectives and Lemma 19 for first-order quantification.

**Lemma 13** *Let $\varphi$ be one of the atomic formulae $\top$, $\bot$, $\lambda(x) = a$ or $x = y$ and let $V$ be a set of variables with $\mathrm{FV}(\varphi) \subseteq V$. Then for all morphisms $h : B^* \to A^*$ and all $\delta : \mathbb{V}_1 \to_p \mathbb{N}$ there exists a formula $h_\delta^{-1}(\varphi)$ which for all fragments $\mathcal{F}$ and all $n \geq \max_{b \in B} |h(b)|$ satisfies*

$$h_\delta^{-1}(\varphi) \leq_\mathcal{F} \varphi \quad \text{and}$$
$$[\![h_\delta^{-1}(\varphi)]\!]_{B,V_n} = h_\delta^{-1}([\![\varphi]\!]_{A,V}).$$

*Proof:* We have $h_\delta(w) \in [\![\top]\!]_{A,V}$ if $\delta(x)$ is defined and if $1 \leq \delta(x) \leq |h_\delta(\lambda_w(x))|$ for all first-order variables $x \in V$. Therefore, if $\delta(x)$ is undefined for some $x \in V \cap \mathbb{V}_1$, then we let $h_\delta^{-1}(\top) = \bot$. Else we let

$$h_\delta^{-1}(\top) = \bigwedge_{x \in V \cap \mathbb{V}_1} \lambda(x) \in B_{\delta(x)}$$

where $B_i = \{b \in B \mid 1 \leq i \leq |h(b)|\}$ is the set of letters $b \in B$ such that $i$ is a position of $h(b)$. By the above considerations we see $[\![h_\delta^{-1}(\top)]\!]_{B,V_n} = h_\delta^{-1}([\![\top]\!]_{A,V})$. Note that if $h$ is length-multiplying, then we could also set $h_\delta^{-1}(\top) = \top$ if $\delta(x)$ is defined and in $\{1, \ldots, m\}$ and $h_\delta^{-1}(\top) = \bot$ otherwise; here $m = |h(b)|$ for some $b \in B$. Consider a position $i$ of $h(w)$ with $h$-coordinates $(j, d)$. Then $\lambda_{h(w)}(i) = a$ is equivalent to $\lambda_{h(b)}(d) = a$ where $b = \lambda_w(j)$. Let $C = \{b \in B \mid \lambda_{h(b)}(\delta(x)) = a\}$ be the set of letters $b \in B$ such that $h(b)$ has label $a$ at position $\delta(x)$. Let $i'$ be a position of $h(w)$ with $h$-coordinates $(j', d')$. Then $i = i'$ if and only if $j = j'$ and $d = d'$. We therefore let

$$\begin{aligned}
h_\delta^{-1}(\bot) &= \bot \\
h_\delta^{-1}(\lambda(x) = a) &= h_\delta^{-1}(\top) \wedge \lambda(x) \in C, \\
h_\delta^{-1}(x = y) &= h_\delta^{-1}(\top) \wedge \begin{cases} x = y & \text{if } \delta(x) = \delta(y), \\ \bot & \text{else.} \end{cases}
\end{aligned}$$

It is easy to see, that these formulae satisfy the syntactic property $h_\delta^{-1}(\varphi) \leq_{\mathcal{F}} \varphi$ for all fragments $\mathcal{F}$. $\square$

**Lemma 14** *Let $x \lesssim y$ be one of the atomic formulae $x < y$ or $x \leq y$ for some $x, y \in \mathbb{V}_1$. Let $V$ be a set of variables with $\mathrm{FV}(x \lesssim y) \subseteq V$. Then for all morphisms $h : B^* \to A^*$ and all $\delta : \mathbb{V}_1 \to_p \mathbb{N}$ there exists a formula $h_\delta^{-1}(x \lesssim y)$ which for all order-stable fragments $\mathcal{F}$ and all $n \geq \max_{b \in B} |h(b)|$ satisfies*

$$\begin{aligned}
h_\delta^{-1}(x \lesssim y) &\leq_{\mathcal{F}} (x \lesssim y) \quad \text{and} \\
[\![h_\delta^{-1}(x \lesssim y)]\!]_{B,V_n} &= h_\delta^{-1}([\![x \lesssim y]\!]_{A,V}).
\end{aligned}$$

*Moreover, if $h$ is length-reducing, then $h_\delta^{-1}(x \lesssim y) \leq_{\mathcal{F}} (x \lesssim y)$ for all fragments $\mathcal{F}$.*

*Proof:* Consider positions $i$ and $i'$ of $h(w)$ with $h$-coordinates $(j, d)$ and $(j', d')$, respectively. Suppose $h$ is length-reducing. Then $d = d' = 1$ and hence $i < i'$ if and only if $j < j'$ and $i \leq i'$ if and only if $j \leq j'$. Thus we let $h_\delta^{-1}(x < y) = (h_\delta^{-1}(\top) \wedge x < y)$ and $h_\delta^{-1}(x \leq y) = (h_\delta^{-1}(\top) \wedge x \leq y)$ where $h_\delta^{-1}(\top)$ is the formula from Lemma 13 for the set $V$. It is easy to see that these formulae satisfies the syntactic property $h_\delta^{-1}(x < y) \leq_{\mathcal{F}} (x < y)$ and $h_\delta^{-1}(x \leq y) \leq_{\mathcal{F}} (x \leq y)$ for all fragments $\mathcal{F}$.

Suppose now $h$ is not length-reducing. Then $i < i'$ if $j < j'$ or if $j \leq j'$ and $d < d'$; and $i \leq i'$ if $j < j'$ or if $j \leq j'$ and $d \leq d'$. Let therefore

$$\begin{aligned}
h_\delta^{-1}(x < y) &= h_\delta^{-1}(\top) \wedge \begin{cases} x \leq y & \text{if } \delta(x) < \delta(y), \\ x < y & \text{else,} \end{cases} \\
h_\delta^{-1}(x \leq y) &= h_\delta^{-1}(\top) \wedge \begin{cases} x \leq y & \text{if } \delta(x) \leq \delta(y), \\ x < y & \text{else.} \end{cases}
\end{aligned}$$

It is easy to verify that these formulae satisfies the syntactic property $h_\delta^{-1}(x < y) \leq_{\mathcal{F}} (x < y)$ and $h_\delta^{-1}(x \leq y) \leq_{\mathcal{F}} (x \leq y)$ for all order-stable fragments $\mathcal{F}$. $\square$

**Lemma 15** *Let $\varphi$ be one of the atomic formulae $\mathrm{suc}(x,y)$, $\min(x)$, $\max(x)$ or empty and let $V$ be a set of variables with $\mathrm{FV}(\varphi) \subseteq V$. Then for all non-erasing morphism $h : B^* \to A^*$ and all $\delta : \mathbb{V}_1 \to_p \mathbb{N}$ there exists a formula $h_\delta^{-1}(\varphi)$ which for all fragments $\mathcal{F}$ and all $n \geq \max_{b \in B} |h(b)|$ satisfies*

$$h_\delta^{-1}(\varphi) \leq_\mathcal{F} \varphi \quad \text{and}$$
$$[\![h_\delta^{-1}(\varphi)]\!]_{B,V_n} = h_\delta^{-1}([\![\varphi]\!]_{A,V}).$$

*Proof:* Suppose $h$ is non-erasing. Clearly, $h(w)$ is empty only if $w$ is empty. Consider a position $i$ of $h(w)$ with $h$-coordinates $(j, d)$. Then $i = 1$ is equivalent to $j = d = 1$; note that for non-erasing morphisms we may have $i = 1$ but nonetheless $j > 1$. We therefore let

$$\begin{aligned}
h_\delta^{-1}(\text{empty}) &= \text{empty}, \\
h_\delta^{-1}(\min(x)) &= \begin{cases} h_\delta^{-1}(\top) \wedge \min(x) & \text{if } \delta(x) = 1, \\ \bot & \text{else.} \end{cases}
\end{aligned}$$

For the max predicate we observe that $i = |h(w)|$ if $j = |w|$ and $\delta = |h(b)|$ where $b = \lambda_w(j)$; note that for erasing morphisms we may have $i = |h(w)|$ but nonetheless $j < |w|$. For suc consider positions $i$ and $i'$ of $h(w)$ with $h$-coordinates $(j, d)$ and $(j', d')$, respectively. Suppose $i + 1 = i'$. We consider two cases. If $d' = 1$, then necessarily $j + 1 = j'$ and $d = |h(b)|$ where $b = \lambda_w(j)$; otherwise $j = j'$ and $d + 1 = d'$. Note that for erasing morphisms we may have $i + 1 = i$ but $j + 1 < j$. Now, let $C = \{b \in B \mid \delta(x) = |h(b)|\}$ be the set of labels, for which $\delta(x)$ is the maximum position in the image under $h$. Then we let

$$\begin{aligned}
h_\delta^{-1}(\max(x)) &= h_\delta^{-1}(\top) \wedge \lambda(x) \in C \wedge \max(x) \\
h_\delta^{-1}(\mathrm{suc}(x,y)) &= \begin{cases} h_\delta^{-1}(\top) \wedge \mathrm{suc}(x,y) \wedge \lambda(x) \in C & \text{if } \delta(y) = 1, \\ h_\delta^{-1}(\top) \wedge x = y & \text{else if } \delta(x) + 1 = \delta(y), \\ \bot & \text{else.} \end{cases}
\end{aligned}$$

It is easy to see, that these formulae satisfy the syntactic property $h_\delta^{-1}(\varphi) \leq_\mathcal{F} \varphi$ for all fragments $\mathcal{F}$. $\square$

**Lemma 16** *Consider the atomic formula $x \equiv r \pmod{q}$ for some $x \in \mathbb{V}_1$ and some $r, q \in \mathbb{Z}$. Let $\mathfrak{F}$ be a family of fragments $\mathcal{F}$ such that $x \equiv r \pmod{q}$ and $x \equiv s \pmod{q}$ are $\mathcal{F}$-equivalent for all $s \in \mathbb{Z}$. Let $V$ be a set of variables with $\mathrm{FV}(x \equiv r \pmod{q}) \subseteq V$. Then for all length-multiplying morphisms $h : B^* \to A^*$ and all $\delta : \mathbb{V}_1 \to_p \mathbb{N}$ there exists a formula $h_\delta^{-1}(x \equiv r \pmod{q})$ which for all $\mathcal{F} \in \mathfrak{F}$ and all $n \geq |h(b)|$ for $b \in B$ satisfies*

$$h_\delta^{-1}\big(x \equiv r \pmod{q}\big) \leq_\mathcal{F} \big(x \equiv r \pmod{q}\big) \quad \text{and}$$
$$[\![h_\delta^{-1}\big(x \equiv r \pmod{q}\big)]\!]_{B,V_n} = h_\delta^{-1}\big([\![x \equiv r \pmod{q}]\!]_{A,V}\big).$$

*Moreover, if $h$ is length-preserving, then $h_\delta^{-1}\big(x \equiv r \pmod{q}\big) \leq_\mathcal{F} \big(x \equiv r \pmod{q}\big)$ for all fragments $\mathcal{F}$.*

*Proof:* Let $h$ be length-multiplying and let $m = |h(b)|$ for some $b \in B$. If $m = 0$ or if $\delta(y)$ is undefined for some $y \in V \cap \mathbb{V}_1$, then $h_\delta^{-1}(x \equiv r \pmod{q}) = \bot$. Let $i$ be a position of $h(w)$ with $h$-coordinates $(j, d)$. Since $h$ is length-multiplying, we have $i = m(j - 1) + d$. Let $t = \gcd(q, m)$ be the greatest common divisor of $q$ and $m$. Let $q = pt$, $m = \ell t$ and let $r' = r + m - d$. Then $i \equiv r \pmod{q}$ if and only if $mj \equiv r' \pmod{q}$. Hence, $t$ is a divisor of $r'$ and $mj \equiv r' \pmod{q}$ is equivalent to $\ell j \equiv r'/t \pmod{p}$. Since $\gcd(\ell, p) = 1$ there exists a number $\ell^{-1}$ such that $\ell^{-1}\ell \equiv 1 \pmod{p}$ and $\ell j \equiv r'/t \pmod{p}$ if and only if $j \equiv \ell^{-1}r'/t \pmod{p}$. Now the latter is equivalent to the existence of $0 \le k < t$ such that $j \equiv \ell^{-1}r'/t + kp \pmod{q}$.

These considerations lead to the following formula. If $r + m - \delta(x) \not\equiv 0 \pmod{t}$, then let $h_\delta^{-1}(x \equiv r \pmod{q}) = \bot$. Let otherwise $s = \ell^{-1}(r + m - \delta(x))/t$ and

$$h_\delta^{-1}(x \equiv r \pmod{q}) \;=\; h_\delta^{-1}(\top) \wedge \bigvee_{k=0}^{t-1} x \equiv s + kp \pmod{q}.$$

It is easy to see the syntactic property $h_\delta^{-1}(x \equiv r \pmod{q}) \le_\mathcal{F} (x \equiv r \pmod{q})$ for all fragments $\mathcal{F} \in \mathfrak{F}$.

Note that if $h$ is length-preserving, then $h_\delta^{-1}(x \equiv r \pmod{q}) = h_\delta^{-1}(\top) \wedge x \equiv r \pmod{q}$ because $m = 1$, $t = 1$ and $s = r$. In this case $h_\delta^{-1}(x \equiv r \pmod{q}) \le_\mathcal{F} (x \equiv r \pmod{q})$ for all fragments $\mathcal{F}$. $\square$

**Lemma 17** *Consider the atomic formula $x \in X$ for some $x \in \mathbb{V}_1$ and $X \in \mathbb{V}_2$. Let $V$ be a set of variables with $\mathrm{FV}(x \in X) \subseteq V$. Then for all morphisms $h : B^* \to A^*$ and all $\delta : \mathbb{V}_1 \to_p \mathbb{N}$ there exists a formula $h_\delta^{-1}(x \in X)$ which for all MSO-stable fragments $\mathcal{F}$ and all $n \ge \max_{b \in B} |h(b)|$ satisfies*

$$h_\delta^{-1}(x \in X) \le_\mathcal{F} (x \in X) \quad \text{and}$$
$$[\![h_\delta^{-1}(x \in X)]\!]_{B, V_n} = h_\delta^{-1}([\![x \in X]\!]_{A, V}).$$

*Moreover, if $h$ is length-reducing, then $h_\delta^{-1}(x \in X) \le_\mathcal{F} (x \in X)$ for all fragments $\mathcal{F}$.*

*Proof:* Let $i$ be a position of $h(w)$ with $h$-coordinates $(j, d)$. Then, by definition of $h_\delta$, we see $\lambda_{h_\delta(w)}(i) = (a, J)$ for some $J \subseteq V$ with $X \in J$ if and only if $\lambda_w(j) = (b, J')$ for some $J' \subseteq V_n$ with $X_d \in J'$. Now, if $\delta(x) \notin \{1, \ldots, N\}$ where $N = \max_{b \in B} |h(b)|$, then $h_\delta^{-1}(x \in X) = \bot$; otherwise let

$$h_\delta^{-1}(x \in X) \;=\; h_\delta^{-1}(\top) \wedge x \in X_{\delta(x)}.$$

This formula is easily seen to satisfy the syntactic property $h_\delta^{-1}(x \in X) \le_\mathcal{F} (x \in X)$ for all MSO-stable fragments $\mathcal{F}$. Moreover, if $h$ is length-reducing, then clearly $h_\delta^{-1}(x \in X) \le_\mathcal{F} (x \in X)$ for all fragments $\mathcal{F}$; notice that $X_1 = X$ by definition of $V_n$. $\square$

Next, we give the following "lifting lemmas": If for some formulae the inverse morphic images are definable, then so are the inverse morphic images of their Boolean combinations (Lemma 18), their first-order and second-order quantification (Lemma 19 and Lemma 20, respectively) and their modular counting quantification (Lemma 21). Moreover, the construction respects every family of morphisms and every collection of fragments (in the case of second-order quantification every collection of MSO-stable fragments; and for the modular counting quantifier every collection of mod-stable fragments).

**Lemma 18** *Let $\psi$ be one of the formulae $\neg \varphi_1$ or $\varphi_1 \vee \varphi_2$ or $\varphi_1 \wedge \varphi_2$. Let $\mathfrak{F}$ be a family of fragments and let $\mathcal{C}$ be a family of morphisms between finitely generated free monoids. Suppose for all sets of variables $V \supseteq \mathrm{FV}(\varphi_i)$, all $\mathcal{C}$-morphisms $h : B^* \to A^*$ and all $\delta : \mathbb{V}_1 \to_p \mathbb{N}$ there exist formulae $h_\delta^{-1}(\varphi_i)$, $i \in \{1, 2\}$, which for all $\mathcal{F} \in \mathfrak{F}$ and all $n \geq \max_{b \in B} |h(b)|$ satisfy*

$$h_\delta^{-1}(\varphi_i) \leq_\mathcal{F} \varphi_i \quad \text{and}$$
$$[\![h_\delta^{-1}(\varphi_i)]\!]_{B,V_n} = h_\delta^{-1}([\![\varphi_i]\!]_{A,V}).$$

*Let $V$ be a set of variables with $\mathrm{FV}(\psi) \subseteq V$. Then for all $\mathcal{C}$-morphisms $h : B^* \to A^*$ and all $\delta : \mathbb{V}_1 \to_p \mathbb{N}$ there exists a formula $h_\delta^{-1}(\psi)$ which for all $\mathcal{F} \in \mathfrak{F}$ and all $n \geq \max_{b \in B} |h(b)|$ satisfies*

$$h_\delta^{-1}(\psi) \leq_\mathcal{F} \psi \quad \text{and}$$
$$[\![h_\delta^{-1}(\psi)]\!]_{B,V_n} = h_\delta^{-1}([\![\psi]\!]_{A,V}).$$

*Proof:* The formulae for disjunction and conjunction are straightforward:

$$h_\delta^{-1}(\varphi_1 \vee \varphi_2) = h_\delta^{-1}(\varphi_1) \vee h_\delta^{-1}(\varphi_2),$$
$$h_\delta^{-1}(\varphi_1 \wedge \varphi_2) = h_\delta^{-1}(\varphi_1) \wedge h_\delta^{-1}(\varphi_2).$$

For the negation, we have to take more care. For $w \in [\![\top]\!]_{B,V_n}$ we may have $h_\delta(w) \notin [\![\varphi_1]\!]_{A,V}$ simply because $h_\delta(w) \notin [\![\top]\!]_{A,V}$. This may happen, e.g., if $\delta$ is such that $h_\delta(w)$ does not allow to interpret all first-order variables of $V$. Therefore $[\![\neg h_\delta^{-1}(\varphi_1)]\!]_{B,V_n}$ is not a subset of $[\![\top]\!]_{B,V_n}$ in general. This is enforced by a conjunction with $h_\delta^{-1}(\top)$ and we let

$$h_\delta^{-1}(\neg \varphi_1) = h_\delta^{-1}(\top) \wedge \neg h_\delta^{-1}(\varphi_1).$$

It is easy to verify that the syntactic property of the $h_\delta^{-1}(\varphi_i)$ conveys to $h_\delta^{-1}(\psi)$, i.e., we have $h_\delta^{-1}(\psi) \leq_\mathcal{F} \psi$ for all $\mathcal{F} \in \mathfrak{F}$. Note that $\neg h_\delta^{-1}(\varphi_1) \leq_\mathcal{F} \neg \varphi_1$. □

**Lemma 19** *Consider the formulae $\exists x\, \varphi$ and $\forall x\, \varphi$ for some $x \in \mathbb{V}_1$. Let $\mathfrak{F}$ be a family of fragments and let $\mathcal{C}$ be a family of morphisms between finitely generated free monoids. Suppose for all sets of variables $V \supseteq \mathrm{FV}(\varphi_i)$, for all $\mathcal{C}$-morphisms $h : B^* \to A^*$ and all $\delta : \mathbb{V}_1 \to_p \mathbb{N}$ there exists a formula $h_\delta^{-1}(\varphi)$ which for all $\mathcal{F} \in \mathfrak{F}$ and all $n \geq \max_{b \in B} |h(b)|$ satisfies*

$$h_\delta^{-1}(\varphi) \leq_\mathcal{F} \varphi \quad \text{and}$$
$$[\![h_\delta^{-1}(\varphi)]\!]_{B,V_n} = h_\delta^{-1}([\![\varphi]\!]_{A,V}).$$

*Let $V$ be a set of variables with $\mathrm{FV}(\exists x\, \varphi) = \mathrm{FV}(\forall x\, \varphi) \subseteq V$. Then for all $\mathcal{C}$-morphisms $h : B^* \to A^*$ and all $\delta : \mathbb{V}_1 \to_p \mathbb{N}$ there exist formulae $h_\delta^{-1}(\exists x\, \varphi)$ and $h_\delta^{-1}(\forall x\, \varphi)$ which for all $\mathcal{F} \in \mathfrak{F}$ and all $n \geq \max_{b \in B} |h(b)|$ satisfy the following properties:*

$$h_\delta^{-1}(\exists x\, \varphi) \leq_\mathcal{F} \exists x\, \varphi, \qquad [\![h_\delta^{-1}(\exists x\, \varphi)]\!]_{B,V_n} = h_\delta^{-1}([\![\exists x\, \varphi]\!]_{A,V}),$$
$$h_\delta^{-1}(\forall x\, \varphi) \leq_\mathcal{F} \forall x\, \varphi, \qquad [\![h_\delta^{-1}(\forall x\, \varphi)]\!]_{B,V_n} = h_\delta^{-1}([\![\forall x\, \varphi]\!]_{A,V}).$$

*Proof:* Let $i$ be a position of $h_\delta(w)$ with $h$-coordinates $(j,d)$. Then $h_\delta(w)[x/i] = h_{\delta[x/d]}(w[x/j])$ where $\delta[x/d]$ is given by $x \mapsto d$ and $y \mapsto \delta(y)$ if $y \neq x$. This leads to

$$h_\delta^{-1}(\exists x\ \varphi) = \exists x \bigvee_{d=1}^{N} h_{\delta[x/d]}^{-1}(\varphi)$$

$$h_\delta^{-1}(\forall x\ \varphi) = \forall x \bigwedge_{d=1}^{N} h_{\delta[x/d]}^{-1}(\varphi)$$

where $N = \max_{b \in B} |h(b)|$ and $h_{\delta[x/d]}^{-1}(\varphi)$ is the formula from the premise for the mapping $h_{\delta[x/d]}$ and set of variables $V \cup \{x\}$. Note that for the $h$-coordinates $(j,d)$ of every position of $h(w)$ we have $1 \le d \le N$ by choice of $N$. The syntactic properties $h_\delta^{-1}(\exists x\ \varphi) \le_\mathcal{F} \exists x\ \varphi$ and $h_\delta^{-1}(\forall x\ \varphi) \le_\mathcal{F} \forall x\ \varphi$ for all $\mathcal{F} \in \mathfrak{F}$ are easily verified. □

**Lemma 20** *Consider the formulae $\exists X\ \varphi$ and $\forall X\ \varphi$ for some $X \in \mathbb{V}_2$. Let $\mathfrak{F}$ be a family of fragments and let $\mathcal{C}$ be a family of morphisms between finitely generated free monoids. Suppose for all sets of variables $V \supseteq \mathrm{FV}(\varphi)$, for all $\mathcal{C}$-morphisms $h: B^* \to A^*$ and all $\delta: \mathbb{V}_1 \to_p \mathbb{N}$ there exists a formula $h_\delta^{-1}(\varphi)$ which for all $\mathcal{F} \in \mathfrak{F}$ and all $n \ge \max_{b \in B} |h(b)|$ satisfies*

$$h_\delta^{-1}(\varphi) \le_\mathcal{F} \varphi \quad \text{and}$$
$$[\![h_\delta^{-1}(\varphi)]\!]_{B,V_n} = h_\delta^{-1}([\![\varphi]\!]_{A,V}).$$

*Let $V$ be a set of variables with $\mathrm{FV}(\exists X\ \varphi) = \mathrm{FV}(\forall X\ \varphi) \subseteq V$. Then for all $\mathcal{C}$-morphisms $h: B^* \to A^*$ and all $\delta: \mathbb{V}_1 \to_p \mathbb{N}$ there exist formulae $h_\delta^{-1}(\exists X\ \varphi)$ and $h_\delta^{-1}(\exists X\ \varphi)$ which for all MSO-stable fragments $\mathcal{F} \in \mathfrak{F}$ and all $n \ge \max_{b \in B} |h(b)|$ satisfy the following properties:*

$$h_\delta^{-1}(\exists X\ \varphi) \le_\mathcal{F} \exists X\ \varphi, \quad [\![h_\delta^{-1}(\exists X\ \varphi)]\!]_{B,V_n} = h_\delta^{-1}([\![\exists X\ \varphi]\!]_{A,V}),$$
$$h_\delta^{-1}(\forall X\ \varphi) \le_\mathcal{F} \forall X\ \varphi, \quad [\![h_\delta^{-1}(\forall X\ \varphi)]\!]_{B,V_n} = h_\delta^{-1}([\![\forall X\ \varphi]\!]_{A,V}).$$

*Moreover, if $h$ is length-reducing, then $h_\delta^{-1}(\exists X\ \varphi) \le_\mathcal{F} \exists X\ \varphi$ and $h_\delta^{-1}(\forall X\ \varphi) \le_\mathcal{F} \forall X\ \varphi$ for all fragments $\mathcal{F} \in \mathfrak{F}$.*

*Proof:* Let $i$ be a position of $h_\delta(w)$ with $h$-coordinates $(j,d)$. Then $\lambda_{h_\delta(w)}(i) \in A \times (\{X\} \cup 2^V)$ if $\lambda_w(j) \in B \times (\{X_d\} \cup 2^{V_n})$ by definition of $h_\delta$. Hence we let

$$h_\delta^{-1}(\exists X\ \varphi) = \exists X_1 \cdots \exists X_N\ h_\delta^{-1}(\varphi),$$
$$h_\delta^{-1}(\forall X\ \varphi) = \forall X_1 \cdots \forall X_N\ h_\delta^{-1}(\varphi)$$

where $N = \max(\{1\} \cup \{|h(b)| \mid b \in B\})$ and $h_\delta^{-1}(\varphi)$ is the formula from the premise for the set of variables $V \cup \{X\}$. Note that for the $h$-coordinates $(j,d)$ of every position of $h(w)$ we have $1 \le d \le N$ by choice of $N$. It is easily verified that the syntactic properties $h_\delta^{-1}(\exists X\ \varphi) \le_\mathcal{F} \exists X\ \varphi$ and $h_\delta^{-1}(\forall X\ \varphi) \le_\mathcal{F} \forall X\ \varphi$ hold for $\mathcal{F} \in \mathfrak{F}$ if $\mathcal{F}$ is MSO-stable or if $h$ is length-reducing. Notice that $X_1 = X$ by definition of $V_n$, and that if $h$ is length-reducing, then $N = 1$. □

The following lemma lifts the inverse morphism construction to modular counting quantifiers. It in particular applies to mod-stable fragments, but holds in a more general setting. Specifically, we do not need the closure under negation required for mod-stability.

**Lemma 21** *Consider the formula $\exists^{r \bmod q} x\, \varphi$ for some $x \in \mathbb{V}_1$ and some $r, q \in \mathbb{Z}$. Let $\mathfrak{F}$ be a family of fragments and let $\mathcal{C}$ be a family of morphisms between finitely generated free monoids. Suppose for all sets of variables $V \supseteq \mathrm{FV}(\varphi)$, for all $\mathcal{C}$-morphisms $h : B^* \to A^*$ and all $\delta : \mathbb{V}_1 \to_p \mathbb{N}$ there exists a formula $h_\delta^{-1}(\varphi)$ which for all $\mathcal{F} \in \mathfrak{F}$ and all $n \geq \max_{b \in B} |h(b)|$ satisfies*

$$h_\delta^{-1}(\varphi) \leq_\mathcal{F} \varphi \quad \text{and}$$
$$[\![h_\delta^{-1}(\varphi)]\!]_{B,V_n} = h_\delta^{-1}([\![\varphi]\!]_{A,V}).$$

*Let $V$ be a set of variables with $\mathrm{FV}(\exists^{r \bmod q} x\, \varphi) \subseteq V$. Then for all $\mathcal{C}$-morphisms $h : B^* \to A^*$ and all $\delta : \mathbb{V}_1 \to_p \mathbb{N}$ there exists a formula $h_\delta^{-1}(\exists^{r \bmod q} x\, \varphi)$ which for all $\mathcal{F} \in \mathfrak{F}$ with $\left(\exists^{r \bmod q} x\, \varphi\right) \equiv_\mathcal{F} \left(\exists^{s \bmod q} x\, \varphi\right)$ (for all $s \in \mathbb{Z}$) and all $n \geq \max_{b \in B} |h(b)|$ satisfies*

$$h_\delta^{-1}(\exists^{r \bmod q} x\, \varphi) \leq_\mathcal{F} \exists^{r \bmod q} x\, \varphi \quad \text{and}$$
$$[\![h_\delta^{-1}(\exists^{r \bmod q} x\, \varphi)]\!]_{B,V_n} = h_\delta^{-1}([\![\exists^{r \bmod q} x\, \varphi]\!]_{A,V}).$$

*Moreover, if $h$ is length-reducing, then $h_\delta^{-1}(\exists^{r \bmod q} x\, \varphi) \leq_\mathcal{F} \exists^{r \bmod q} x\, \varphi$ for all $\mathcal{F} \in \mathfrak{F}$.*

*Proof:* For $d \in \mathbb{N}$ by $\delta[x/d] : \mathbb{V}_1 \to_p \mathbb{N}$ we denote the mapping $x \mapsto d$ and $y \mapsto \delta(y)$ if $y \neq x$. Let $h_{\delta[x/d]}^{-1}(\varphi)$ be the formula from the premise for the set of variables $V \cup \{x\}$. Let

$$h_\delta^{-1}(\exists^{r \bmod q} x\, \varphi) \;=\; \bigvee_{s \in S} \bigwedge_{d=1}^{N} \exists^{s(d) \bmod q} x\; h_{\delta[x/d]}^{-1}(\varphi)$$

where $N = \max_{b \in B} |h(b)|$ and $S$ is the set of functions $s : \{1, \ldots, N\} \to \{0, \ldots, q-1\}$ such that $\sum_{d=1}^N s(d) \equiv r \pmod{q}$. A position is a $\varphi$-*position* if $\varphi$ holds when $x$ is interpreted to be this position. The idea is that the $\varphi$-positions of $h_\delta(w)$ are partitioned; for every $d \in \{1, \ldots, N\}$ the number of $\varphi$-positions of $h_\delta(w)$ originating from a position in $w$ with offset $d$ is counted separately (modulo $q$). The total sum of these counts then has to be $r$ (modulo $q$). Note that $S$ is finite and that for the $h$-coordinates $(j, d)$ of every position of $h(w)$ we have $1 \leq d \leq N$ by choice of $N$. Hence, every $\varphi$-position of $h_\delta(w)$ is counted in precisely one of the terms of the conjunction. The syntactic property $h_\delta^{-1}(\exists^{r \bmod q} x\, \varphi) \leq_\mathcal{F} \exists^{r \bmod q} x\, \varphi$ for all $\mathcal{F} \in \mathfrak{F}$ with $\left(\exists^{r \bmod q} x\, \varphi\right) \equiv_\mathcal{F} \left(\exists^{s \bmod q} x\, \varphi\right)$ is easily verified.

Suppose $h$ is length-reducing, i.e., $N \leq 1$. Consider first the case $N = 0$. If $r \equiv 0 \pmod{q}$, then $h_\delta^{-1}(\exists^{r \bmod q} x\, \varphi) = \top$ and else $h_\delta^{-1}(\exists^{r \bmod q} x\, \varphi) = \bot$. If $N = 1$, then $S$ contains only the function $s$ with $s(1) = (r \bmod q)$ and we redefine $h_\delta^{-1}(\exists^{r \bmod q} x\, \varphi) = \exists^{r \bmod q} x\; h_{\delta[x/1]}^{-1}(\varphi)$. In both cases the formulae satisfy $h_\delta^{-1}(\exists^{r \bmod q} x\, \varphi) \leq_\mathcal{F} \exists^{r \bmod q} x\, \varphi$ for all $\mathcal{F} \in \mathfrak{F}$. □

We are now ready to prove Proposition 2.

**Proposition 2.** *Let $\mathcal{F}$ be a fragment and let $\mathcal{C}$ be a family of morphisms between finitely generated free monoids. Suppose the following:*

1. *If $\mathcal{F}$ contains a second-order quantifier, then $\mathcal{F}$ is MSO-stable or all $\mathcal{C}$-morphisms are length-reducing.*

2. If $\mathcal{F}$ contains the predicate $\leq$ or $<$, then $\mathcal{F}$ is order-stable or all $\mathcal{C}$-morphisms are length-reducing.

3. If $\mathcal{F}$ contains the predicate suc, min, max or empty, then all $\mathcal{C}$-morphisms are non-erasing.

4. If $\mathcal{F}$ contains a modular predicate, then all $\mathcal{C}$-morphisms are length-multiplying and either $\mathcal{F}$ is mod-stable or all $\mathcal{C}$-morphisms are length-preserving.

5. If $\mathcal{F}$ contains a modular quantifier, then $\mathcal{F}$ is mod-stable or all $\mathcal{C}$-morphisms are length-reducing.

Then the class of languages defined by $\mathcal{F}$ is closed under inverse $\mathcal{C}$-morphisms.

*Proof:* We show by induction on the structure of $\varphi$ that for all sets of variables $V$ containing $\mathrm{FV}(\varphi)$ for all $\mathcal{C}$-morphisms $h : B^* \to A^*$ and for all $\delta : \mathbb{V}_1 \to_p \mathbb{N}$ there exists a formula $h_\delta^{-1}(\varphi)$ which satisfies $h_\delta^{-1}(\varphi) \leq_\mathcal{F} \varphi$ and $[\![h_\delta^{-1}(\varphi)]\!]_{B,V_n} = h_\delta^{-1}([\![\varphi]\!]_{A,V})$ for all $n \geq \max_{b \in B} |h(b)|$. For the atomic modalities $\top$, $\bot$, $\lambda(x) = a$ and $x = y$ this is Lemma 13; for $x < y$ and $x \leq y$ it is Lemma 14; for $\mathrm{suc}(x,y)$, $\min(x)$, $\max(x)$ and empty it is Lemma 15; for the modular predicate $x \equiv r \pmod{q}$ it is Lemma 16; and for $x \in X$ it is Lemma 17. Note that in all cases the lemmas do apply. For Boolean connectives, first-order quantification, second-order quantification and modular quantification, this follows by induction and Lemma 18, Lemma 19, Lemma 20 and Lemma 21, respectively (where $\mathfrak{F} = \{\mathcal{F}\}$). With this claim closure under inverse morphic images follows readily: Suppose $\varphi$ is a sentences and let $h^{-1}(\varphi) = h_\delta^{-1}(\varphi)$ for some arbitrary $\delta$. Then we get $h^{-1}(L_A(\varphi)) = L_B(h^{-1}(\varphi))$ and $h^{-1}(\varphi) \leq_\mathcal{F} \varphi$. In particular $\varphi \in \mathcal{F}$ implies $h^{-1}(\varphi) \in \mathcal{F}$, that is, if $L \in \mathcal{L}_A(\mathcal{F})$, then $h^{-1}(L) \in \mathcal{L}_B(\mathcal{F})$. □

# E. Remaining Proofs from Section 3

**Corollary 1.** *Let $\mathcal{F} \subseteq \mathrm{MSO}[<,\leq,=]$ be a fragment which is MSO-stable and order-stable. Then $\mathcal{F}$ defines a positive $*$-variety.*

*Proof:* This is a direct consequence of Theorem 1. □

**Corollary 2.** *Let $\mathcal{F} \subseteq \mathrm{MSO}[<,\leq,=,\mathrm{suc},\min,\max]$ be a fragment which is MSO-stable and order-stable. Suppose $\min(y) \leq_\mathcal{F} \mathrm{suc}(x,y)$ and $\max(x) \leq_\mathcal{F} \mathrm{suc}(x,y)$ for all first-order variables $x$ and $y$. Then the languages defined by $\mathcal{F}$ over nonempty words is a positive $+$-variety.*

*Proof:* Let $L = L_A(\varphi) \cap A^+$ be the language defined by $\varphi \in \mathcal{F}_A$ over $A^+$. We first show closure under left residuals. Let $\mathcal{G}$ be the smallest fragment containing $\mathcal{F}$ and satisfying empty $\leq_\mathcal{G} \min(x)$ and empty $\leq_\mathcal{G} \max(x)$ for all first-order variables $x$. Then $\mathcal{G}$ is suc-stable and MSO-stable. Of course, $L_A(\varphi) \in \mathcal{L}_A(\mathcal{G})$ and hence $a^{-1}L_A(\varphi) \in \mathcal{L}_A(\mathcal{G})$ via some formula $a^{-1}\varphi \in \mathcal{G}$ because $\mathcal{L}(G)$ is closed under residuals by Proposition 1. Replacing in $a^{-1}\varphi$ each predicate empty by $\bot$ to obtain $\psi$, we get $\psi \leq_\mathcal{F} a^{-1}\varphi$. Hence $\psi \in \mathcal{G}$ which implies $\psi \in \mathcal{F}$. But $\psi$ and $a^{-1}\varphi$ are equivalent over nonempty words, i.e, $L_A(\psi) \cap A^+ = L_A(a^{-1}\varphi) \cap A^+ = a^{-1}L_A(\varphi) \cap A^+ = a^{-1}L \cap A^+$. This shows that $\psi$ defines the left residual by $a$ of $L$ over $A^+$.

Closure under right residuals follows by symmetry and closure of $\mathcal{L}(\mathcal{F})$ under inverse non-erasing morphisms is an immediate consequence of Proposition 2. □

# F. Proof of Theorem 2

**Theorem 2.** *A language is definable in* $\mathrm{FO}^2[<]$ *if and only if it is definable by a formula in* $\Sigma_2[<, \leq]$ *with an acyclic comparison graph.*

*Proof:* Suppose $L \subseteq A^*$ is $\mathrm{FO}^2[<]$-definable. Then $L$ is a finite union of *unambiguous monomials*, i.e., languages of the form $P = A_1^* a_1 \cdots A_n^* a_n A_{n+1}^*$ such that each $w \in P$ has a unique factorization $w = w_1 a_1 \cdots w_n a_n w_{n+1}$ with $w_i \in A_i^*$; see [20, 5]. The parameter $n$ is the *degree* of $P$. There exists $i \in \{1, \ldots, n\}$ such that $a_i \notin A_1 \cap A_{n+1}$ because otherwise $(a_1 \cdots a_n)^2$ admits two different factorizations. Suppose $a_i \notin A_1$ and let $a = a_i$. Making the first occurrence of $a$ explicit in $P$ shows that $P$ is a finite union of languages of the form $Q_1 a Q_2$ where $Q_1$ and $Q_2$ are unambiguous monomials with degree smaller than $k$. Inductively, there exist $\varphi_j \in \Sigma_2[<, \leq]$ with acyclic comparison graph such that $\varphi_j$ defines $Q_j$. We may assume that the variables used by $\varphi_1$ and $\varphi_2$ are disjoint. The next step is to relativize $\varphi_1$ to the factor to the left of the first $a$-position. For this let

$$\varphi_1' = \exists x_1 \exists x_2 \left( \varphi_1(<) \wedge \bigwedge_{j \in \{1,2\}} \lambda(x_j) = a \wedge \forall y \, (x_j \leq y \vee \lambda(y) \neq a) \right) \quad (1)$$

where $x_1, x_2, y$ are new variables. Before turning to the formula $\varphi_1(<)$ note that both $x_1$ and $x_2$ specify the first $a$-position thus ensuring $x_1 = x_2$ without actually using equality. (This will become important for acyclicity.) Also take notice that $\lambda(y) \neq a$ can be expressed positively by $\lambda(y) \in A \setminus \{a\}$. The construction of $\varphi_1(<)$ is by induction on the structure of the formula. Let $\psi(<) = \psi$ for atomic formulae, $(\neg \psi)(<) = \neg \psi(<)$, $(\psi_1 \, \text{⋈} \, \psi_2)(<) = \psi_1(<) \, \text{⋈} \, \psi_2(<)$ for $\text{⋈} \in \{\vee, \wedge\}$ and

$$\begin{aligned} (\exists z \, \psi)(<) &= \exists z \, (z < x_1 \wedge \psi(<)), \\ (\forall z \, \psi)(<) &= \forall z \, (x_2 \leq z \vee \psi(<)). \end{aligned}$$

The formula $\varphi_1'$ holds on a word if and only if the word has an $a$-position and $\varphi_1$ holds on the factor before the first $a$-position. One can verify that $\varphi_1'$ is acyclic since $\varphi_1$ is. A similar construction yields an acyclic formula $\varphi_2''$ which evaluates $\varphi_2$ on the factor beyond the first $a$-position. This shows that $Q_1 a Q_2$ is defined by the acyclic formula $\varphi_1' \wedge \varphi_2''$. Hence, $P$ is a disjunction of acyclic formulae which, after renaming variables, yields an acyclic formula. The construction for $a_i \notin A_{n+1}$ is similar but the $x_j$ in (1) then specify the last $a$-position, i.e., "$x_j \leq y$" is replaced by "$y \leq x_j$".

Let $L \subseteq A^*$ be defined by $\varphi \in \Sigma_2[<, \leq]$ with $G(\varphi) = (V, E)$ acyclic. Suppose $\varphi$ is in prenex form, i.e., $\varphi = \exists x_1 \cdots \exists x_k \forall y_{k+1} \cdots \forall y_\ell \, \psi$ where $\psi$ is quantifier-free. We shall show that $p(uv)^n u(uv)^n q \in L \Leftrightarrow p(uv)^{3n} q \in L$ for $n \geq \ell^2$ and $u, v, p, q \in A^*$. From this it follows that the syntactic monoid of $L$ is in **DA** which is known to be equivalent to $L$ being definable in $\mathrm{FO}^2[<]$, see [20, 5].

The implication $p(uv)^{3n} q \in L(\varphi) \Rightarrow p(uv)^n u(uv)^n q \in L(\varphi)$ holds for all $\varphi \in \Sigma_2[<, \leq]$ without acyclicity condition [20, 5]. It therefore suffices to show the converse implication. We may assume that $E$ is a linear order on $V$ such that $(j, j+1) \in E$ for $1 \leq j < k$ and for $k < j < \ell$. For simplicity, we identify variables with their interpretation on a word. Consider an interpretation $x_1, \ldots, x_k \in \mathbb{N}$ on $p(uv)^n u(uv)^n q$ such that for all interpretations of the $y_j$ the formula $\psi$ holds on $p(uv)^n u(uv)^n q$. By choice of $n$, there exists a factorization $p(uv)^n u(uv)^n q = p'(uv)^\ell w(uv)^\ell q'$ with $p'(uv)^\ell$ being a prefix of $p(uv)^n$ and $(uv)^\ell q'$ being a suffix of $(uv)^n q$ and such that none of the $x_j$ is in one of the factors $(uv)^\ell$ of this factorization. Specifically, $x_j \in I_1 \cup I_2 \cup I_3$ for $1 \leq j \leq k$ where

- $I_1 = \{1, \ldots, |p'|\}$,
- $I_2 = \{|p'(uv)^\ell| + 1, \ldots, |p'(uv)^\ell w|\}$, and
- $I_3 = \{|p'(uv)^\ell w(uv)^\ell| + 1, \ldots, |p(uv)^n u(uv)^n q|\}$.

Let $p(uv)^{3n}q = p'(uv)^\ell w'(uv)^\ell q'$ and let $I_1'$, $I_2'$, and $I_3'$ be defined analogously to $I_1$, $I_2$ and $I_3$, respectively, with $w$ replaced by $w'$. Let $\pi : \mathbb{N} \to \mathbb{N}$ be an order-respecting injection mapping $I_j$ to $I_j'$, $j \in \{1, 2, 3\}$.

We construct an interpretation $x_j'$ of the $x_j$ on $p(uv)^{3n}q$ as follows. For $x_j \in I_1 \cup I_3$ we set $x_j' = \pi(x_j)$. For the positions $x_j \in I_2$ we let $x_j' \in I_2'$ such that $x_j'$ and $x_j$ have the same label and such that $j < j'$ implies $x_j' < x_{j'}'$. Assume there exists an interpretation $y_{k+1}', \ldots, y_\ell'$ on $p(uv)^{3n}q$ for $y_{k+1}, \ldots, y_\ell$ making $\psi$ false. We are going to construct an interpretation of the $y_j$ on $p(uv)^n u(uv)^n q$ such that $\psi$ is false. If $y_j' \in I_1' \cup I_3'$, then $y_j = \pi^{-1}(y_j')$. Every $y_j' \notin I_1' \cup I_3'$ is classified into "left" or "right" as follows. If $y_j' < \min\{y_{j'}' \in I_2' \mid j' \leq k\}$, then $y_j'$ is "left". If $y_j' > \max\{y_{j'}' \in I_2' \mid j' \leq k\}$, then $y_j'$ is "right". Else $y_{j'}' \leq y_j' \leq y_{j''}'$ for some $j', j'' \leq k$. We distinguish three cases:

1. If $(j', j) \in E$ and $(j'', j) \in E$, then $y_j'$ is "left".
2. If $(j, j') \in E$ and $(j, j'') \in E$, then $y_j'$ is "right".
3. If $(j', j) \in E$ and $(j, j'') \in E$, then $y_j'$ is "right".

(In the third case the classification does not really matter.) Note that by acyclicity, $(j'', j) \in E$ and $(j, j') \in E$ cannot happen. The "left" (respectively "right") positions are set label-respecting in the range between $|p'| + 1$ and $|p'(uv)^\ell|$ (between $|p'(uv)^\ell w| + 1$ and $|p'(uv)^\ell w(uv)^\ell|$, respectively) such that $y_j < y_{j'}$ for "left"-positions (respectively "right"-positions) $y_j'$ and $y_{j'}'$ with $j' < j$. By construction every atomic formula which is true on $p(uv)^n u(uv)^n q$ is also true on $p(uv)^{3n}q$. Since $\psi$ does not contain negations, it is monotonic in its atoms and thus $\psi$ is false on $p(uv)^n u(uv)^n q$ for this valuation, a contradiction. $\square$